\documentclass{emulateapj}

\usepackage{natbib}

\usepackage{graphicx}
\usepackage{latexsym}
\usepackage{amsfonts,amsmath,amssymb}
\usepackage{url}
\usepackage[utf8]{inputenc}
\usepackage{fancyref}
\usepackage{hyperref}

\begin{document}

\title{Numerical simulations of winds driven by radiation force from the corona above a thin disk}

\author{Xiao-Hong Yang\altaffilmark{1}, De-Fu Bu\altaffilmark{2} and Qi-Xiu Li \altaffilmark{1}}

\altaffiltext{1}{Department of Physics, Chongqing University, Chongqing 400044, China; yangxh@cqu.edu.cn} \altaffiltext{2}{Shanghai Astronomical Observatory, Chinese Academy of Sciences, 80 Nandan Road, Shanghai 200030, China; dfbu@shao.ac.cn}

\def\LSUN{\rm L_{\odot}}
\def\MSUN{\rm M_{\odot}}
\def\RSUN{\rm R_{\odot}}
\def\MSUNYR{\rm M_{\odot}\,yr^{-1}}
\def\MSUNS{\rm M_{\odot}\,s^{-1}}
\def\MDOT{\dot{M}}

\begin{abstract}
Observations show that winds can be driven from the innermost region (inside a 50 Schwarschild radius) of a thin disk. It is interesting to study the winds launched from the innermost region. A hot corona above the black hole (BH) thin disk is irradiated by the disk. We perform two-dimensional hydrodynamical simulations to study the winds driven by radiation force from the corona in the innermost regions. The hard X-ray spectrum from active galactic nuclei (AGNs) suggests that the corona temperature is about $10^9$ K, so that we mainly analyze the properties of winds (or outflows) from the $10^9$ K corona. The disk luminosity plays an important role in driving the outflows. The more luminous the disk, the stronger the outflows. Mass outflow rate ($\dot{M}_{\rm out}$) at a 90 Schwarschild radius depends on disk luminosity, which can be described as $\dot{M}_{\rm out}\propto 10^{3.3 \Gamma}$ ($\Gamma$ is the ratio of the disk luminosity to the Eddington luminosity). In the case of high luminosity (e.g. $\Gamma=0.75$), the supersonic outflows with maximum speed $1.0 \times 10^4$ Km s$^{-1}$ are launched at $\sim17^{o}$ --$30^{o}$ and $\sim50^{o}$ --$80^{o}$ away from the pole axis. The Bernoulli parameter keeps increasing with the outward propagation of outflows. The radiation force keeps accelerating the outflows when outflows move outward. Therefore, we can expect the outflows to escape from the BH gravity and go to the galactic scale. The interaction between outflows and interstellar medium may be an important AGN feedback process.
\end{abstract}

\keywords{accretion, accretion disk -- black hole physics -- hydrodynamics -- methods: numerical}

\section{Introduction}
Outflows/winds are common in different types of sources, such as active galactic nuclei (AGNs), X-ray binaries, and young stellar objects. For example, high-velocity outflows are observed in many luminous Type 1.0--1.9 AGNs (Gofford et al. 2013). X-ray band spectroscopy study has recently indicated that blueshifted X-ray absorption lines exist in a lot of AGNs (Tombesi et al. 2010, Gofford et al. 2013). Such blueshifted X-ray absorption lines are interpreted as Fe XXV and Fe XXVI K-shell resonance lines produced in highly ionized gas with a mildly relativistic outflow velocity, in the range $v\simeq0.03$--$0.4c$, where $c$ is the light speed. Luminous AGNs are commonly considered to be powered by an optically thick and geometrically thin disk (SSD; Shakura \& Sunyaev 1973) around black holes (BHs). Understanding the production of outflows from the thin disk is an important task for researchers. Previous works suggest magnetic fields, the radiation force, and thermal expansion as three key mechanisms of driving disk winds.

Magnetic fields play an important role in the accretion process through magnetorotational instability (MRI) (Balbus \& Hawley 1991) and they are an important factor that can drive astrophysical jets and disk winds (Blandford \& Payne 1982; Sakurai 1987). In the model suggested by Blandford \& Payne (1982), the magnetic field lines are anchored in a rotating thin disk, the field lines corotate with the disk and then the magneto-centrifugal force can drive winds from the thin disk. The inclination angle of the field lines at the disk surface has a strong influence on driving the magneto-centrifugal winds. When the field lines have a low inclination angle with respect to the disk surface, the disk winds are fed directly from the cold disk surface; however, when the field lines are steep, a hot corona must be present to feed the disk winds via thermal expansion (Cao \& Spruit 1994). Numerical simulations have also studied the Blandford \& Payne mechanism (e.g. Krasnopolsky, Li \& Blandford 1999; 2003; Kato, Kudoh \& Shibata 2002; Porth \& Fendt 2010).

The radiation force is suggested to drive disk winds from a thin disk by theoretical works and numerical simulations (e.g. Bisnovatyi-Kogan \& Blinnikov 1977; Icke 1980; Tajima \& Fukue 1996, 1998; Fukue 1996; Proga et al. 2000; Hirai \& Fukue 2001; Fukue 2004; Cao 2012; 2014). The radiation force exerted on the gas above the thin disk depends on the ionization state of gas. If gas is highly ionized, the radiation force is due to Compton scattering; if gas is weakly ionized, the radiation force is mainly produced through spectral line absorption (line force; Castor et al. 1975). Weakly ionized gas can absorb ultraviolet photons produced from the thin disk through the bound--bound transition and then the line force is exerted on the gas. Based on particle dynamics (the highly ionized gas is considered as electron--positron pair particles), previous works have studied particle winds above the thin disk (Icke 1980; Tajima \& Fukue 1996; 1998), which suggests that the disk radiation plays an important role in driving disk winds. Based on hydrodynamics, previous numerical simulations have found that the radiation force is essential for producing the thin disk winds if the gas thermal energy is low or the magnetic field lines are steep (inclination angle of field lines with respect to disk midplane $>60^0$; Proga 2003). Cao (2012) also pointed out that the radiation force is helpful to magnetically launch the disk winds from the cold disk surface at which the steep field lines are anchored. The radiation force can also directly accelerate the disk winds from the thin disk via line force (Proga et al. 2000; Nomura \& Ohsuga 2017), even with the disk luminosity lower than the Eddington luminosity.

The above mechanisms do not exist alone in astrophysical sources to drive disk winds. Using the analytical method, Fukue (2002) examined a hydrodynamical isothermal wind driven by both the radiation force (due to Compton scattering) and thermal expansion. Later, the effects of magnetic pressure (Fukue 2004) were also taken into account. Cao (2014) investigated the isothermal disk winds driven by magnetic field and radiation force (due to Compton scattering) from a hot corona above a radiation-pressure-dominated thin disk.

Based on numerical simulations, Proga et al. (2000) studied the thin disk winds driven by line force. They found that winds can be driven from 300 to 3000 Schwarzschild radius ($r_{\rm s}$). Nomura \& Ohsuga (2017) performed similar simulations and found that thin disk winds can be launched from 30 to 1500 $r_{\rm s}$ by line force. In addition, Proga (2003) simulated the thin disk winds driven simultaneously by the line force and the magneto-centrifugal force. Line force is very efficient in driving disk winds. However, line force cannot be exerted on the highly ionized gas whose temperature is higher than $10^5$ K. Therefore, it is difficult for line force to drive disk winds from the innermost region (e.g. $r<30 r_{\rm s}$) of a thin disk. There are reasons as follows. First, for example, for a BH of $10^6 M_{\bigodot}$ (with $M_{\bigodot}$ being solar mass), if its surrounding thin disk has luminosity higher than $23\%$ of the Eddington value, the disk effective temperature can be higher than $10^5$ K within 30 $r_{\rm s}$. Second, observations have confirmed a power-law hard X-ray spectrum from luminous AGNs. The hard X-ray spectrum can be reproduced using a $\sim 10^9$ K corona above the thin disk via the inverse Compton scattering (e.g. Liu et al. 2003; Cao 2009; Liu et al. 2016). The strong X-ray radiation can highly ionize the gas in the innermost region (e.g. $r<30 r_{\rm s}$). In summary, line force may be absent for $r<30 r_{\rm s}$.

Observations show that winds can be driven from the innermost region (inside a 50 Schwarschild radius) from a thin disk (e.g. Reeves et al. 2009). In this paper, we study the corona winds from the innermost region of the thin disk driven by radiation force due to Compton scattering. Generally, the corona corotates with the thin disk at the Keplerian or sub-Keplerian speed. The gravitational force can be roughly balanced by the centrifugal force. In this case, even with disk luminosity lower than the Eddington value, radiation force due to Compton scattering may drive the corona gas to form winds. In this paper, we try to simulate disk winds launched from the hot corona above a luminous thin disk by radiation force due to Compton scattering. It is helpful to understand the winds launching from the innermost region of luminous accretion disk, both for supermassive BHs and X-ray binaries. As a first step, in this paper, we neglect the effects of magnetic fields.

We organize our paper as follows: Section 2 introduces our model and method; Section 3 is devoted to analysis of simulations; and Section 4 summarizes and discusses our results.

\section{Model and Method}

\begin{figure}

\scalebox{0.5}[0.5]{\rotatebox{0}{\includegraphics[bb=45 20 500 360]{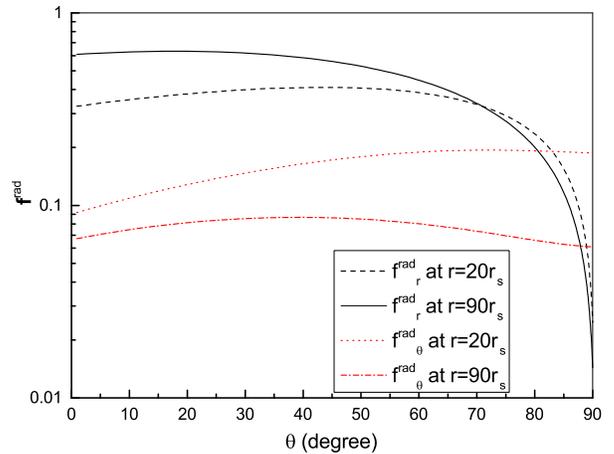}}}
 \ \centering \caption{Angular distribution of the ratio of radiation force to black hole gravity ($\textbf{f}^{\rm rad}=|\textbf{F}^{\rm rad}|/(GM_{\rm bh}/r^2)$) when $\Gamma=0.75$. A black dashed line and a black solid line  correspond to the radial component of radiation force at 20 and 90 $r_{\rm s}$, respectively. A red dotted line and a red dotted-dashed line correspond to the $\theta$ component at 20 and 90 $r_{\rm s}$, respectively.}
 \label{fig 1}
\end{figure}

\begin{table*}
\begin{center}

\caption[]{Summary of models}

\begin{tabular}{cccccc}
\hline\noalign{\smallskip} \hline\noalign{\smallskip}

Run &  $\Gamma$ & $T_{\rm C}$ (K)  &  $\dot{M}_{\rm out}$ ($n\times\dot{M}_{\rm C}$) & $P_{\text{k}}$ ($n\times L_{\rm Edd}$) & $P_{\text{th}}$ ($n\times L_{\rm Edd}$) \\

(1) & (2)             & (3)                         &  (4)      &     (5)   & (6)  \\
\hline\noalign{\smallskip}
R1 & 0.25   &$10^9$        & 9.0$\times10^{-3}$     &3.2$\times10^{-9}$ &2.1$\times10^{-6}$ \\
R2 & 0.50   &$10^9$        & 7.5$\times10^{-2}$     &4.9$\times10^{-7}$ &1.7$\times10^{-5}$ \\
R3 & 0.75   &$10^9$        & 4.2$\times10^{-1}$     &1.2$\times10^{-4}$ &7.1$\times10^{-5}$ \\
T1 & 0.75   &$10^8$        & 6.4$\times10^{-2}$     &1.5$\times10^{-5}$ &6.8$\times10^{-6}$ \\
T2 & 0.75   &$10^7$        & 1.7$\times10^{-2}$     &4.6$\times10^{-6}$ &2.6$\times10^{-6}$ \\
\hline\noalign{\smallskip} \hline\noalign{\smallskip}
\end{tabular}
\end{center}

\begin{list}{}
\item\scriptsize{Column 1: model names; Column 2: the ratio of disk luminosity to Eddington luminosity ($\Gamma=L_{\rm D}/L_{\rm Edd}$); Column 3: the temperature of the hot corona above a thin disk; Columns 4--6: the time-averaged values of the mass outflow rate, and the kinetic and thermal energy fluxes carried out by winds at the outer boundary. The letter ``$n$'' is defined in Equation (10).}
\end{list}
\label{tab1}
\end{table*}

\subsection{Basic Equations}
Shakura \& Sunyaev (1973) proposed the standard thin disk model, which is an optically thick and geometrically thin accretion disk. Observations suggest that there is a hot corona above the thin disk surface. The corona is irradiated by the thin disk. We simulate the irradiated corona using the thin disk. We perform simulations in spherical coordinate ($r$,$\theta$,$\phi$) and assume axis-symmetry. We solve the following hydrodynamical (HD) equations:
\begin{equation}
\frac{\partial\rho}{\partial t}+\nabla\cdot(\rho \mathbf{v})=0,
\label{cont}
\end{equation}
\begin{equation}
\frac{\partial \rho \mathbf{v}}{\partial t}+\nabla\cdot[\rho\mathbf{v}\mathbf{v}+(P)\mathbf{I}]=-\rho\nabla\psi+\rho \mathbf{F}^{\rm rad},
\label{monentum}
\end{equation}
\begin{equation}
\frac{\partial (e+\rho \frac{\mathbf{v}^2}{2})}{\partial t}+\nabla \cdot[(e+\rho \frac{\mathbf{v}^2}{2}+P)\mathbf{v}]=0,
\label{energyequation}
\end{equation}

where $\rho$ , $P$, $\mathbf{v}$, $\psi$ and $e$ are density, pressure, velocity, gravitational potential, and internal energy, respectively. We adopt an equation of state of ideal gas $P=(\gamma -1)e$ and set the adiabatic index $\gamma =5/3$. We also apply the Newtonian potential, $\psi=-GM_{\rm bh}/r$, where $M_{\rm bh}$ and $G$ are the BH mass and the gravitational constant, respectively. $\mathbf{F}^{\rm rad}$ is the total radiation force per unit mass exerted by the thin disk due to Compton scattering.

The hot corona above a thin disk can cool by Compton scattering, thermal conduction, and cycle-synchrotron radiation. The corona can also be heated by some physical mechanisms. The heating mechanism of corona is still an open issue. Liu et al. (2002) suggested magnetic reconnection to heat the corona. Jiang et al. (2014) found dissipation of turbulence driven by MRI to be a possible mechanism. As a first step, for simplicity, we do not take into account these heating and cooling processes. Our assumption is that the heating and cooling processes can balance each other.

Equations \ref{cont}--\ref{energyequation} are numerically solved using the PLUTO code (Mignone et al. 2007; 2012). PLUTO is a Godunov-type code solving the equations of HD and MHD in classical and special relativistic regimes.

\begin{figure}

\scalebox{0.5}[0.64]{\rotatebox{0}{\includegraphics[bb=80 360 490 720]{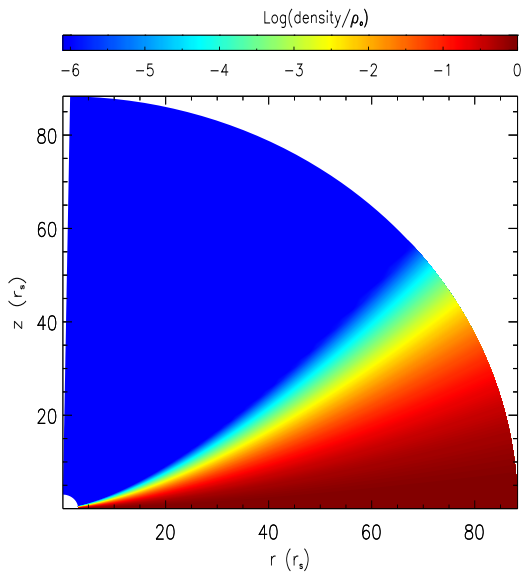}}}
 \ \centering \caption{Initial distribution of normalized density. }
 \label{fig 2}
\end{figure}

\subsection{Radiation Force}
The total luminosity of the standard thin disk is $L_{_{\rm D}}=\Gamma L_{\rm Edd}=GM_{\rm bh}\dot{M}/(2r_{\star})$, where $\Gamma$ is the Eddington ratio of the disk luminosity, $L_{\rm Edd}$ is the Eddington luminosity,  $\dot{M}$ is the accretion rate through the thin disk, and $r_{*}$ is the marginally stable orbit ($\equiv 3 r_{\rm s}= 6GM_{\rm bh}/c^2$). The radiation intensity of the thin disk at the radius $r_{_{_{\rm D}}}$ is given by
\begin{equation}
I_{_{\rm D}}(r_{_{\rm D}})=\frac{3GM_{\rm bh}\dot{M}}{8 \pi^2r^3_{_{\rm D}}}[1-\sqrt{3r_{\rm s}/r_{_{\rm D}}}].
\end{equation}
$I_{_{\rm D}}(r_{_{\rm D}})$ is the frequency-integrated continuum intensity, which is locally isotropic. To calculate the radiation force exerted on the irradiated gas, we have to integrate over the contribution from the whole disk. However, the gravitational instability in the standard thin disk gives rise to fragmentation of the disk beyond 1000$r_{\rm s}$, especially around quasi-stellar objects (Goodman 2003). Then, we integrate over the range of $3r_{\rm s}$--$1000r_{\rm s}$. Due to Compton scattering, the radiation force exerted on the gas element at ($r,\theta$) is given by
\begin{equation}
{\bf F}^{\rm rad}(r,\theta) = \Gamma\frac{6}{\pi} \frac{G M_{\rm bh}}{r^2_{*}}\int^\pi_0 \int^{1000r_{*}/3}_{r_{*}} \hat{\bf n}\frac{r' {\rm cos}(\theta)}{r'^2_{_{\rm D}}d'^{3}_{_{\rm D}}}{\rm d}r'_{_{\rm D}}{\rm d}\phi_{_{\rm D}},
\label{rad_force}
\end{equation}
(see the Appendix of Proga et al. 1998). Here, we express the primed quantities in $r_{*}$ units. ($r_{_{\rm D}}$,$\phi_{_{\rm D}}$) is a polar coordinate located on the disk surface. $\phi_{_{\rm D}}$ is measured from a radiation source on the disk surface to the computational plane ($r,\theta,0^o$). $d_{_{\rm D}}$ is the distance between a radiation source and the irradiated gas element. $\hat{\bf n}(\equiv(n_r,n_\theta,n_{\phi}))$ is the unit vector that directs from the radiation source toward the irradiated gas element. $\hat{\bf n}$ is given by
\begin{equation}
\begin{split}
&n_r=\frac{r-r_{_{\rm D}}{\rm sin}(\theta){\rm cos}(\phi_{_{\rm D}})}{d_{_{\rm D}}},n_\theta=-\frac{r_{_{\rm D}}{\rm cos}(\theta){\rm cos}(\phi_{_{\rm D}})}{d_{_{\rm D}}},\\
&{\rm and}\text{  }n_\phi=-\frac{r_{_{\rm D}}{\rm sin}(\phi_{_{\rm D}})}{d_{_{\rm D}}}.
 \end{split}
\end{equation}
Due to the assumption of axial symmetry, we take into account the radial and $\theta$ components of radiation force. Figure \ref{fig 1} shows angular distribution of the ratio of radiation force to BH gravity ($\textbf{f}^{\rm rad}=|\textbf{F}^{\rm rad}|/(GM_{\rm bh}/r^2)$) at $r=20 \text{ and } 90 r_{s}$ when $\Gamma=0.75$.

Equation \ref{rad_force} assumes that the irradiated gas is optically thin for the disk radiation. This is a reasonable assumption because we take into account the irradiated gas above the photosphere of the disk. The optical depth from infinity to the photosphere should be less than 1. Under this assumption, we can nondimensionalize Equations \ref{cont}--\ref{energyequation} and our results can be used in the arbitrary units system.

\begin{figure}

\scalebox{0.5}[0.5]{\rotatebox{0}{\includegraphics[bb=45 20 500 400]{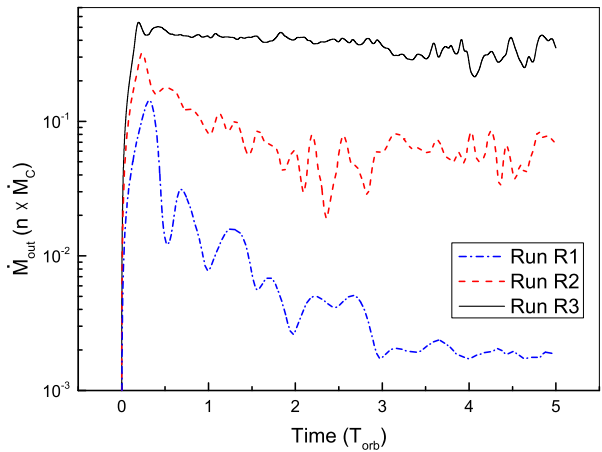}}}
 \ \centering \caption{Time evolution of mass outflow rate at the outer boundary. Different lines correspond to different values of $\Gamma$. The vertical axis is in units of $n\times \dot{M_{\rm}}$, where $n$ is defined in Equation (10).}
 \label{fig 3}
\end{figure}

\begin{figure*}

\scalebox{0.5}[0.64]{\rotatebox{0}{\includegraphics[bb=80 360 520 700]{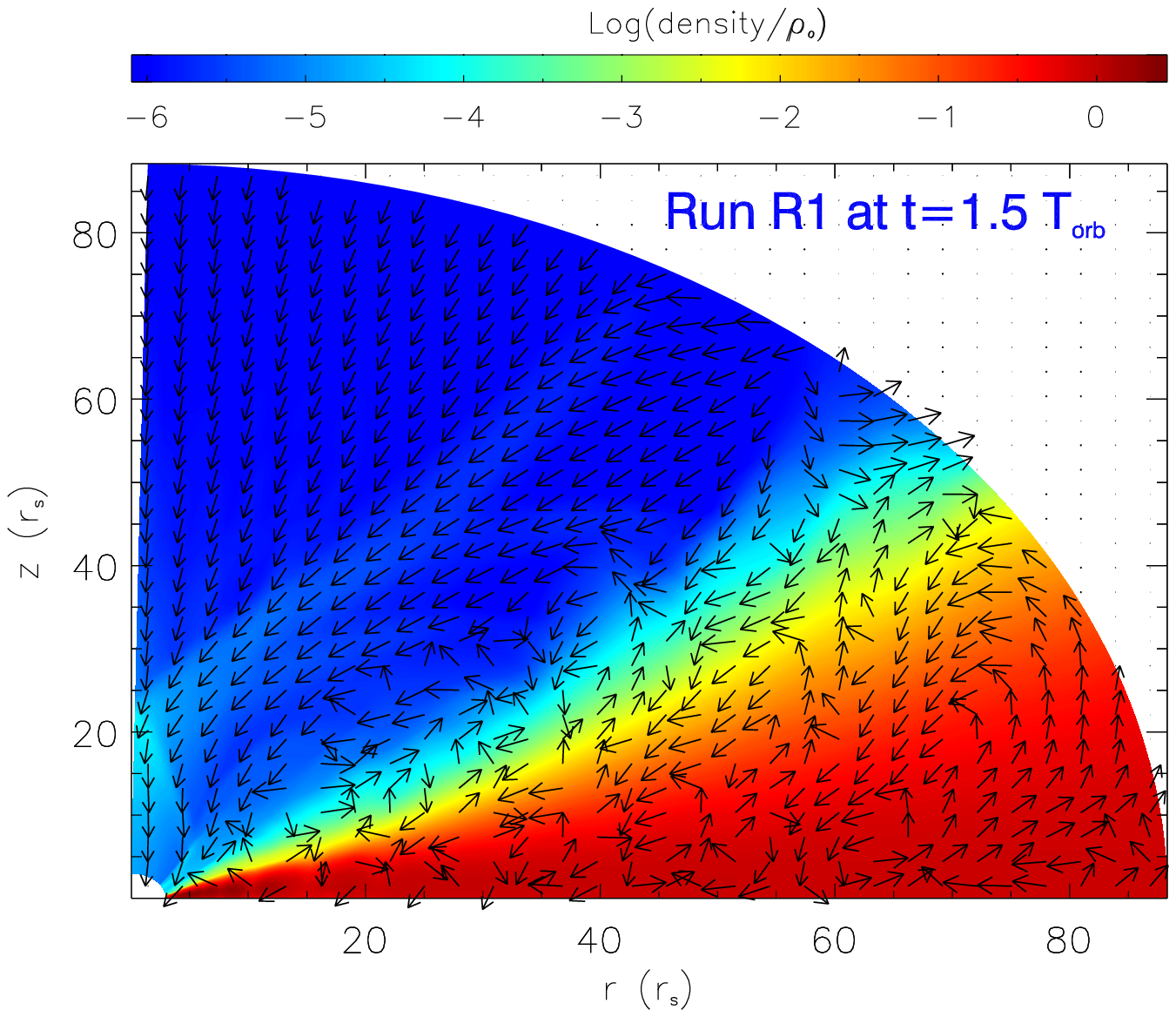}}}
\scalebox{0.5}[0.64]{\rotatebox{0}{\includegraphics[bb=80 360 520 700]{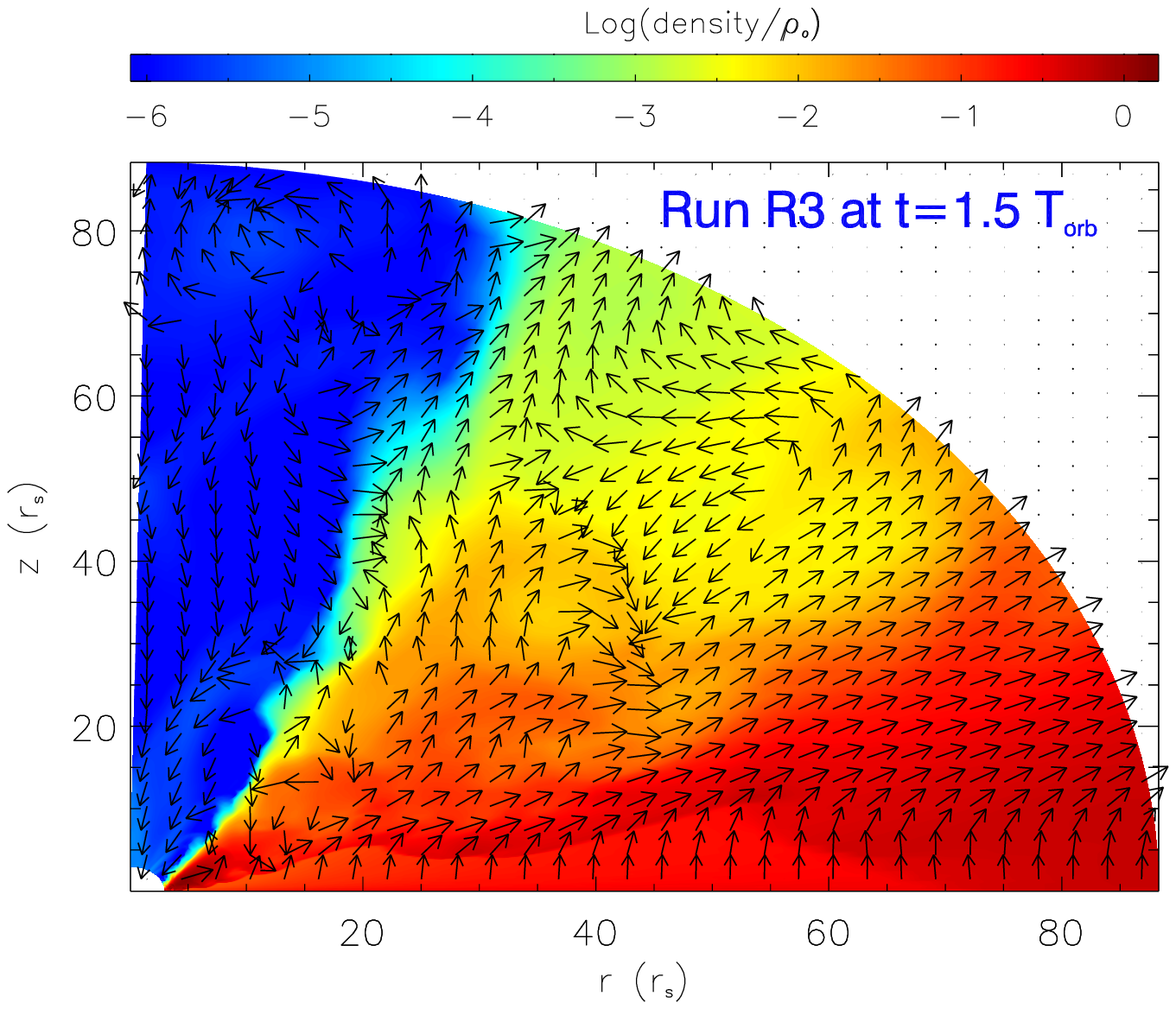}}}
\ \centering \caption{Snapshots of Runs R1 and R3 at $t=1.5 T_{\rm orb}$. Color denotes the logarithm density and arrows denote the direction of the poloidal velocity vector. Left panel: run R1; Right panel: run R3.}
\label{fig 4}
\end{figure*}

\subsection{Model Setup}

Our computational region should be located above the disk surface. Under the simplest thin disk approximation, we make no distinction between the disk surface and the midplane of the thin disk in our simulations. The coordinate origin is set at the black hole of $M_{\rm BH}$. The computational domain covers an angular range of $0 \leq \theta \leq \pi/2$ and a radial range of $r_{\rm in}(=r_{*}) \leq r \leq r_{\rm out}(=30 r_{*})$.  The $\theta=0$ axis is the rotational axis of the thin disk. In the radial direction, we divide the radial range into 144 nonuniform zones in which the zone size ratio is fixed at $(\bigtriangleup r)_{i+1} / (\bigtriangleup r)_{i} = 1.04$. The smallest radial grid zone is $\Delta r =0.0123r_{\rm s}$ at the inner boundary. In the angular direction, we divide the polar angle range into 160 nonuniform zones, i.e. $(\bigtriangleup \theta)_{j+1} / (\bigtriangleup \theta)_{j} = 0.970072$. The smallest angular grid zone is $\Delta \theta =0.023589^{o}$ at the surface of $\theta = \pi/2$.

We set the initial physical variables based on Proga et al. (1998). Under the assumption of isothermal hydrostatic equilibrium in the vertical direction, we have initial density distribution as follows:
\begin{equation}
\rho(r,\theta)=\rho_{_{\rm C}} {\rm exp}(-\frac{GM_{\rm bh}}{2c^2_{\rm s,C}r{\rm tan}^2(\theta)}),
\label{density_distribution}
\end{equation}
where $\rho_{_{\rm C}}$ and $c_{\rm s,C}$ are density and sound speed at the boundary of $\theta=\pi/2$, respectively. Figure \ref{fig 2} shows initial distribution of normalized density. The initial temperature $T(r,\theta)$ is also set to be the temperature ($T_{_{\rm C}}$) of a hot corona above the thin disk. For the initial velocity, ${\bf v}_r(r,\theta)$ and ${\bf v}_{\theta}(r,\theta)$ are set to be null. The rotational velocity is given as ${\bf v}_\phi(r,\theta)={\rm sin}(\theta)\sqrt{GM_{\rm bh}/r}$, which meets the equilibrium between the centrifugal force and the BH gravity. In all of our simulations, we set the density floor to be $10^{-6}\rho_{_{\rm C}}$.

We specify the boundary conditions as follows. At the inner ($r_{\rm in}$) and outer ($r_{\rm out}$) radial boundary, we employ the outflow boundary condition. When the matter goes out of the computational domain, we copy the variable values in the corresponding active zones to the ghost zones. When the matter goes into the computational domain, ${\bf v}_{r}(r,\theta)$ in the ghost zones is set to be zero, while the other variable values are set to the values in the corresponding active zones. At the pole (i.e. $\theta=0$), we apply the axially symmetric boundary condition. At the equator (i.e. $\theta=\pi/2$), we set the fixed values for physical quantities (such as velocity ${\bf v}(r,\pi/2)$, density $\rho(r,\pi/2)$ and temperature $T(r,\pi/2)$) at all times. Density $\rho(r,\pi/2)$ and temperature $T(r,\pi/2)$ are set to the density ($\rho_{\rm C}$) and temperature ($T_{\rm C}$) of a hot corona above the disk surface. For simplicity, $\rho_{\rm C}$ and $T_{\rm C}$ are constant at all radii. ${\bf v}_r(r,\pi/2)$ and ${\bf v}_{\theta}(r,\pi/2)$ is fixed to be null and ${\bf v}_{\phi}(r,\pi/2)$ is fixed to be the Keplerian rotating velocity of the disk. We calculate gas pressure using $P=\frac{\rho k_{\rm B} T}{\mu m_{\rm p}}$, where $\mu$, $m_{\rm p}$, and $k_{_{\rm B}}$ are the mean molecular weight, the proton mass, and the Boltzmann constant, respectively. We set $\mu=0.5$.

\begin{figure*}

\scalebox{0.35}[0.35]{\rotatebox{0}{\includegraphics[bb=45 20 500 360]{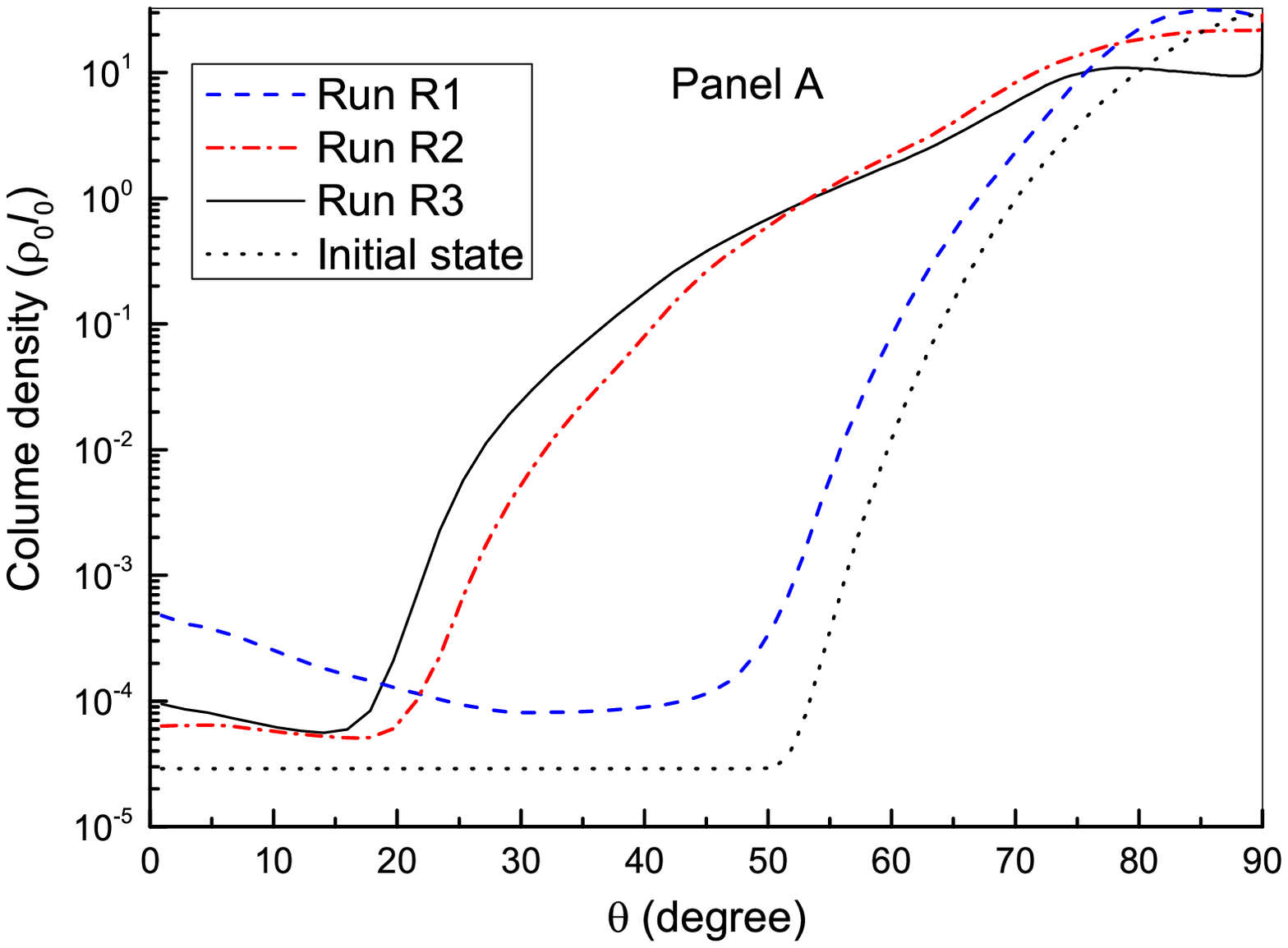}}}
\scalebox{0.35}[0.35]{\rotatebox{0}{\includegraphics[bb=45 20 500 360]{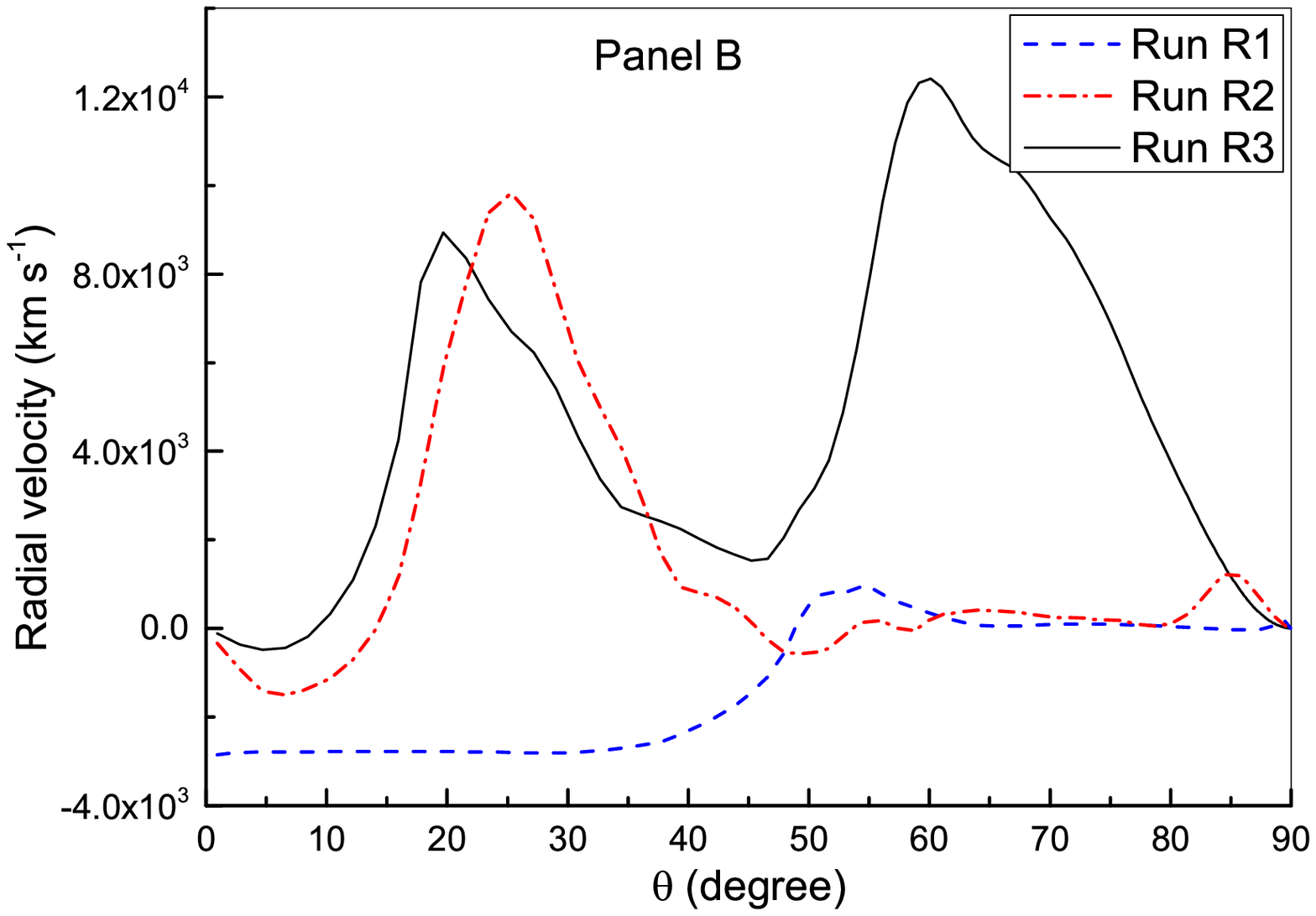}}}
\scalebox{0.35}[0.35]{\rotatebox{0}{\includegraphics[bb=45 20 500 360]{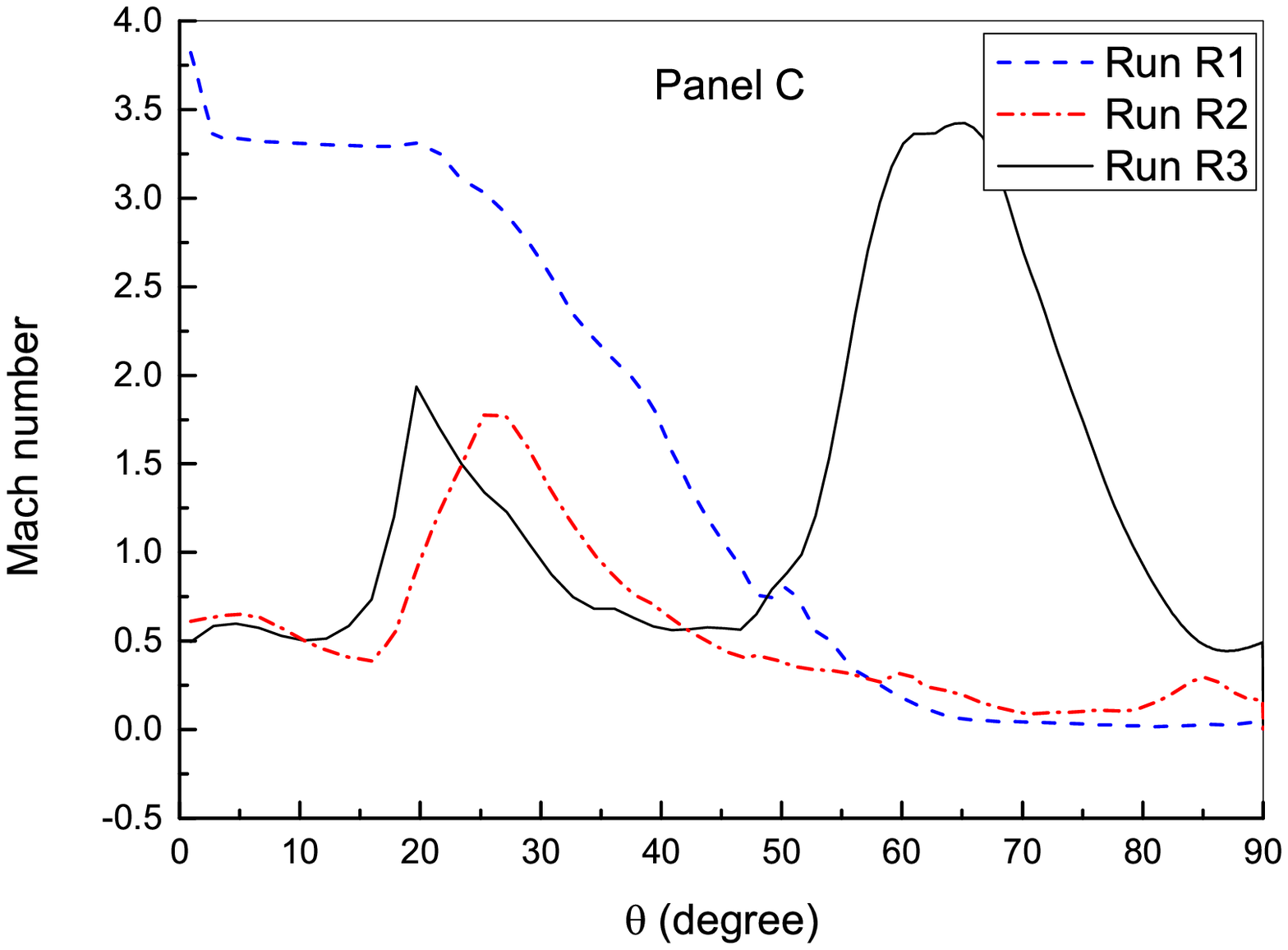}}}
\scalebox{0.35}[0.35]{\rotatebox{0}{\includegraphics[bb=45 20 500 360]{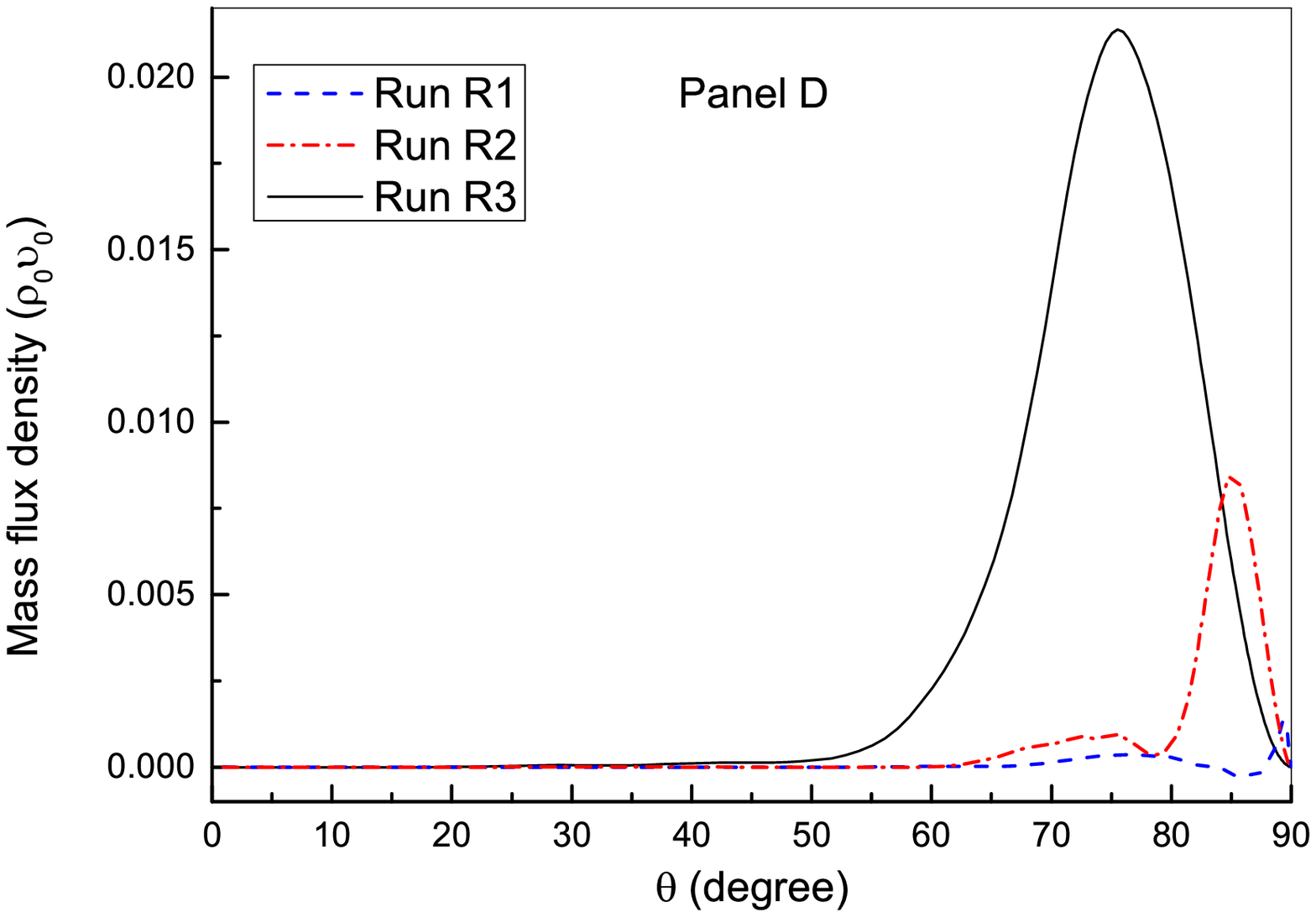}}}
\scalebox{0.35}[0.35]{\rotatebox{0}{\includegraphics[bb=45 20 500 360]{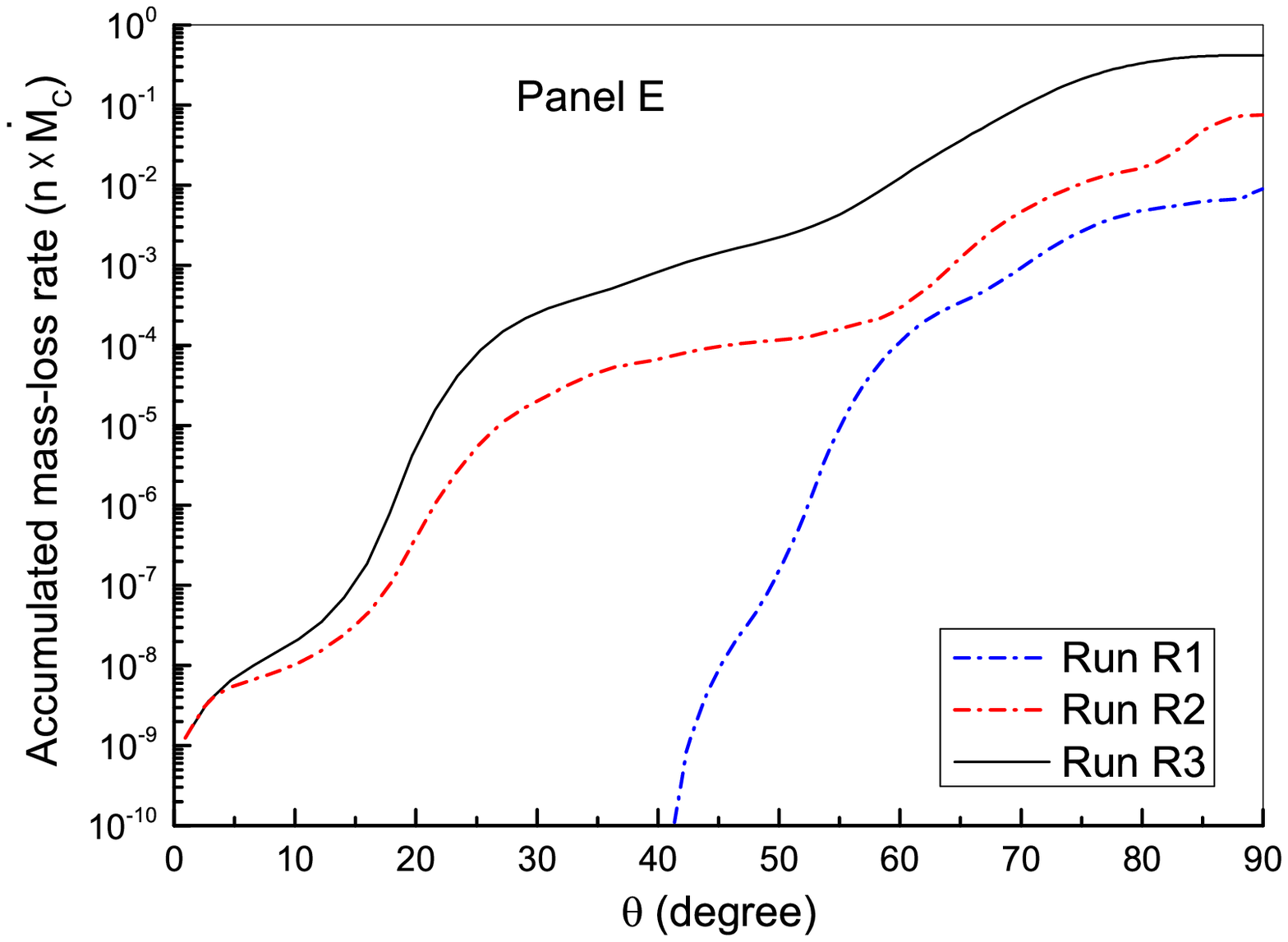}}}

\ \centering \caption{Angular profiles of a variety of time-averaged variables at the outer boundary. Panel (A): column density; panel (B): radial velocity; panel (C): Mach number in the poloidal direction; panel (D): mass flux density; panel (E): accumulated mass outflow rate. The letter ``n'' in panel (D) is defined in Equation (10).}
\label{fig 5}
\end{figure*}

\begin{figure}

\scalebox{0.5}[0.5]{\rotatebox{0}{\includegraphics[bb=45 20 500 360]{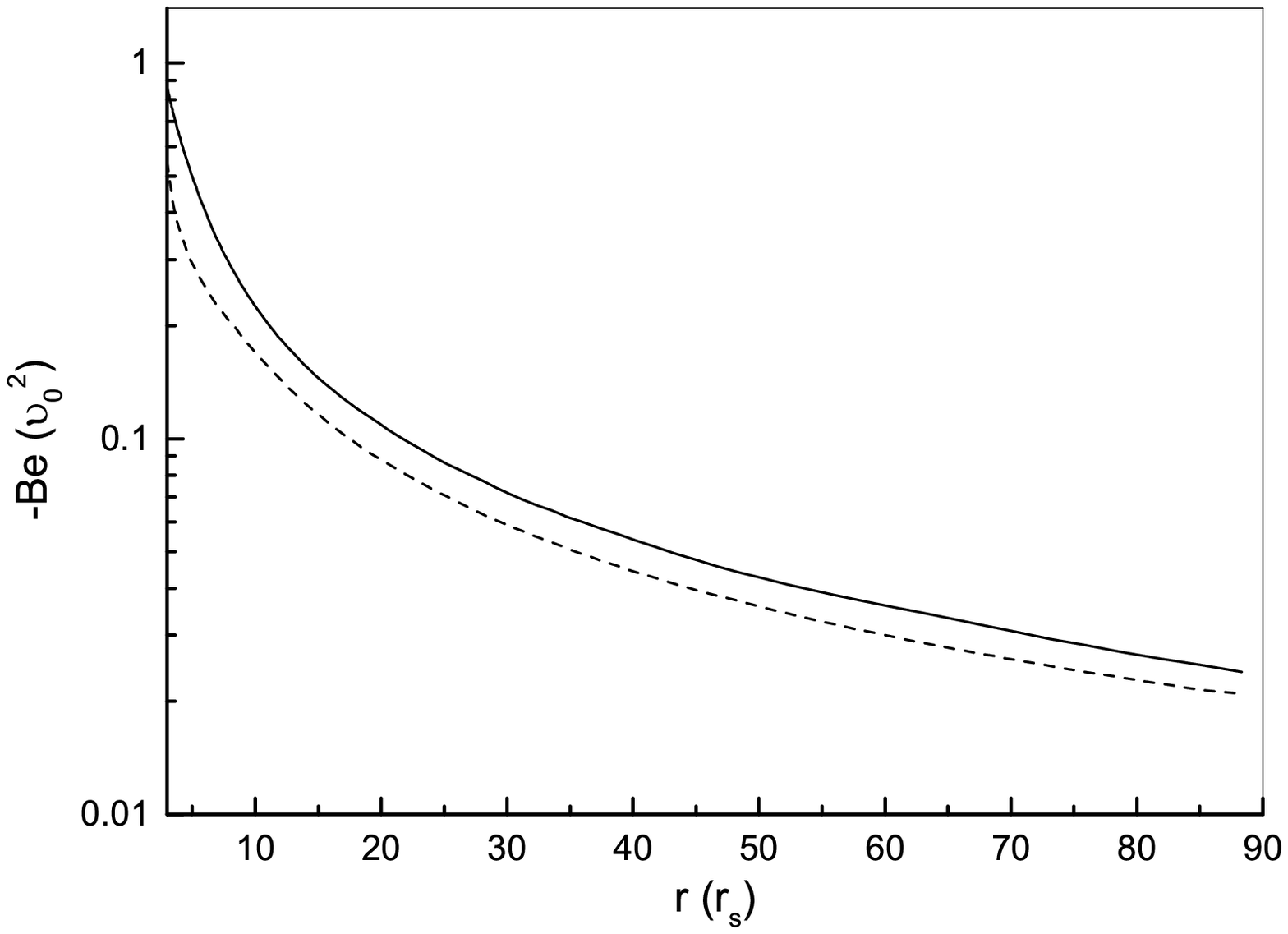}}}
 \ \centering \caption{Radial dependence of the time-averaged Bernoulli parameter ($\text{Be}$) of winds for run R3. The solid line corresponds to an angle-averaged value over an angle between $\theta=10^{o}$ to $\theta=46^{o}$. The dashed line corresponds to an angle-averaged value over an angle between $\theta=46^{o}$ to $\theta=90^{o}$.}
 \label{fig 6_R3}
\end{figure}

\subsection{Units and Model Parameters}
Model parameters are determined by three parameters: the central BH mass ($M_{\rm bh}$), the corona density ($\rho_{_{\rm C}}$), and the corona temperature ($T_{_{\rm C}}$). When we choose a suitable unit system in our simulations, $M_{\rm bh}$ is scale-free. Then, our simulations can be used for various BH systems. In our simulations, the units of length ($l_{0}$), velocity ($\upsilon_{0}$) and density ($\rho_{0}$) are given, respectively, as follows:
\begin{equation}
l_{0}=\frac{6GM_{\rm bh}}{c^2}=8.86\times 10^{13}(\frac{M_{\rm bh}}{10^8 M_{\odot}}) \text{  cm},
\end{equation}
\begin{equation}
\upsilon_{0}=\sqrt{\frac{GM_{\rm bh}}{l_0}}=\frac{c}{\sqrt{6}}=1.22\times 10^{10} \text{  cm s}^{-1},
\end{equation}
\begin{equation}
\rho_{0}=n\times10^{-15}(\frac{M_{\rm bh}}{10^8 M_{\odot}})^{-1} \text{  g cm}^{-3}.
\end{equation}
We can derive the units of mass outflow rate $\dot{m}_{0}=\rho_{0}\upsilon_{0}l_{0}^2$ and then $\dot{m}_{0}= n\times6.87\times10^{-3} \dot{M}_{\rm C}$, where $\dot{M}_{\rm C}=L_{\rm Edd}/c^2$ is the critical mass accretion rate. We also derive the units of energy outflow rate as $\dot{e}_{0}=\rho _{0}\upsilon_{0}^3 l_{0}^2=n\times1.14\times10^{-3} L_{\rm Edd}$.

In this paper, we set the corona density $\rho_{_{\rm C}}=\rho_{0}$ and change the disk luminosity ($\Gamma$) and the corona temperature ($T_{_{\rm C}}$). In our simulations, we do not consider the radiation emission and absorption. For nonradiative simulations, the results are density free. We find that when we apply $\rho_c =\rho_0$ and set $n = 0.1$, the radial Compton scattering optical depth is about 0.1 at the equator. The radial optical depth at the equator always keeps maximum during simulations. If $n<0.1$, the optical depth will be smaller. Therefore, we can believe that our results can be safely applied to the cases of $n<0.1$. We have simulated three cases of the disk luminosity $L_{\rm D}=0.25, 0.5 \text{, and } 0.75 L_{\text{Edd}}$ (i.e. $\Gamma=0.25, 0.5$, and 0.75). Observations have confirmed that the corona temperature is $\sim 10^9$ K (e.g. Liu et al. 2003, 2016; Cao 2009). Therefore, we mainly study models with corona temperature $T_{\rm C}=10^9$ K. For comparison, we have also tested models with lower corona temperature. The model parameters are summarized in Table 1, where we give the mass outflow rate ($\dot{M}_{\rm out}$) as well as the kinetic ($P_{\rm k}$) and thermal energy ($P_{\rm th}$) fluxes carried out by winds at the outer boundary.

\section{Results}

\subsection{Properties of outflows (or winds)}
In our simulations, the centrifugal force initially balances the BH gravity in the horizontal direction. Rotational velocity of gas is set to be Keplerian at the boundary of $\theta=\pi/2$. When the radiation force due to Compton scattering is exerted on the gas, outflows can be launched. In this work, we mainly focus on the properties of outflows (or winds) driven by radiation force. To obtain the time-averaged values of $\dot{M}_{out}$, $P_{\rm k}$ and $P_{\rm th}$ in Table 1, we average the data over the time range of 1.0--2.0 $T_{\rm orb}$, where $T_{\rm orb}$ is the orbital period at the outer boundary (i.e. $90 r_{\rm s}$).

Figure \ref{fig 3} shows the time evolution of mass outflow rate at the outer boundary. The mass outflow rate reaches maximum during $\sim0.3$--0.4$T_{\rm orb}$ and hereafter shows large fluctuation. We take runs R1 and R3 as two examples. Figure \ref{fig 4} shows the density snapshot of runs R1 and R3 at $t=1.5T_{\rm orb}$. We use arrows to denote the direction of poloidal velocity in Figure \ref{fig 4}. As shown in Figure \ref{fig 4}, run R1 ($\Gamma=0.25$) basically maintains the initial density distribution shown in Figure \ref{fig 2}. We have also tested the case of $\Gamma=0.0$ (i.e. no radiation force is exerted on the corona) and found that the steady-state density distribution is very similar to the initial condition shown in Figure \ref{fig 2}. For run R3 with $\Gamma=0.75$, the irradiated corona spreads to a wider angular distribution, compared to the initial distribution of corona. In run R1, gas motion is mussy in corona inside $50 r_{\rm s}$ and the large-scale circulations form outside $50 r_{\rm s}$. In run R3, gas is blown out from the boundary of $\theta=\pi/2$ and the blown-out gas moves upward to form large-scale circulations or outflows. Comparing runs R1 and R3, weak radiation force (such as $\Gamma=0.25$) cannot effectively drive the irradiated gas to form large-scale outflows.

\begin{figure*}

\scalebox{0.35}[0.35]{\rotatebox{0}{\includegraphics[bb=45 20 500 360]{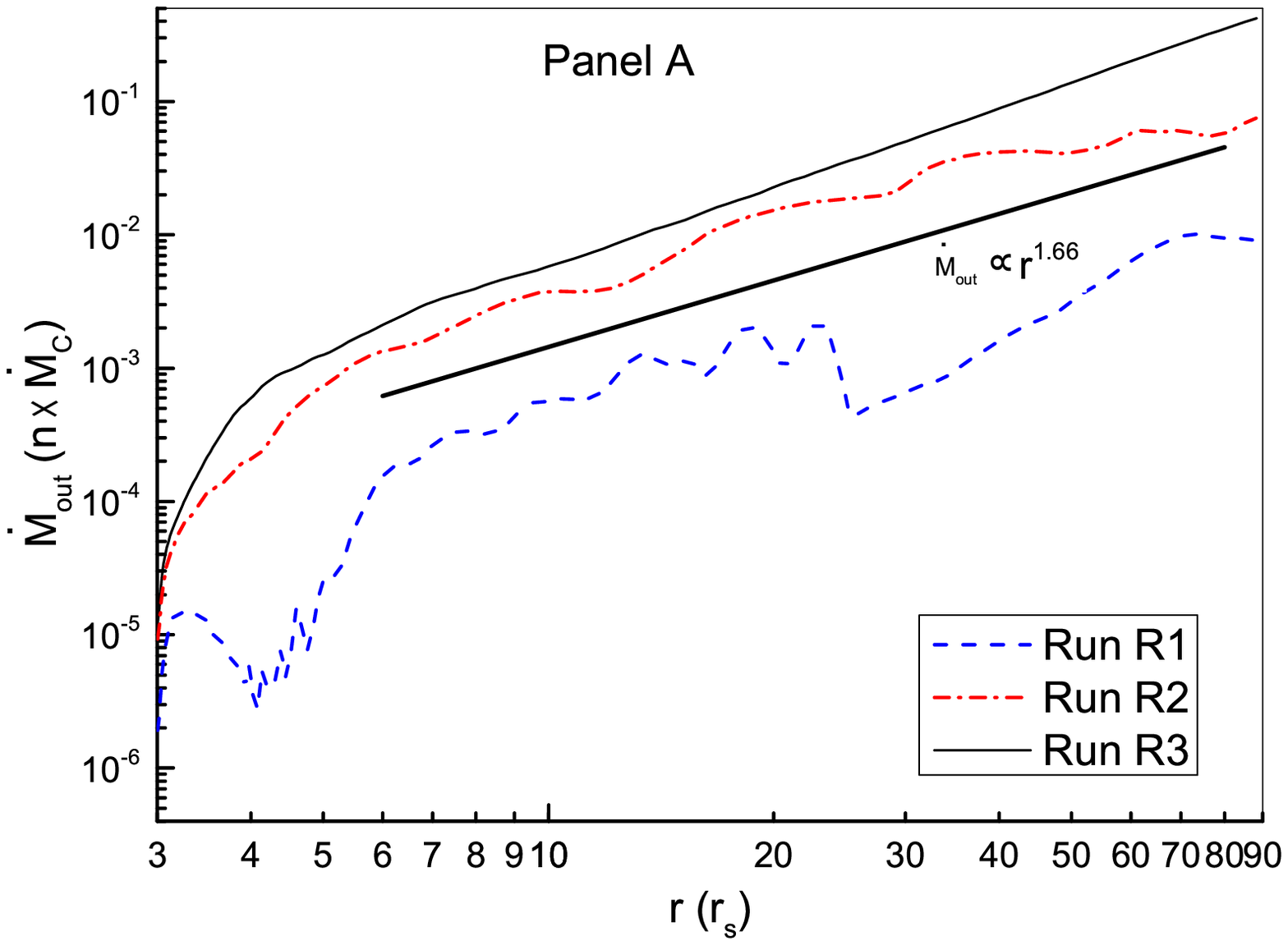}}}
\scalebox{0.35}[0.35]{\rotatebox{0}{\includegraphics[bb=45 20 500 360]{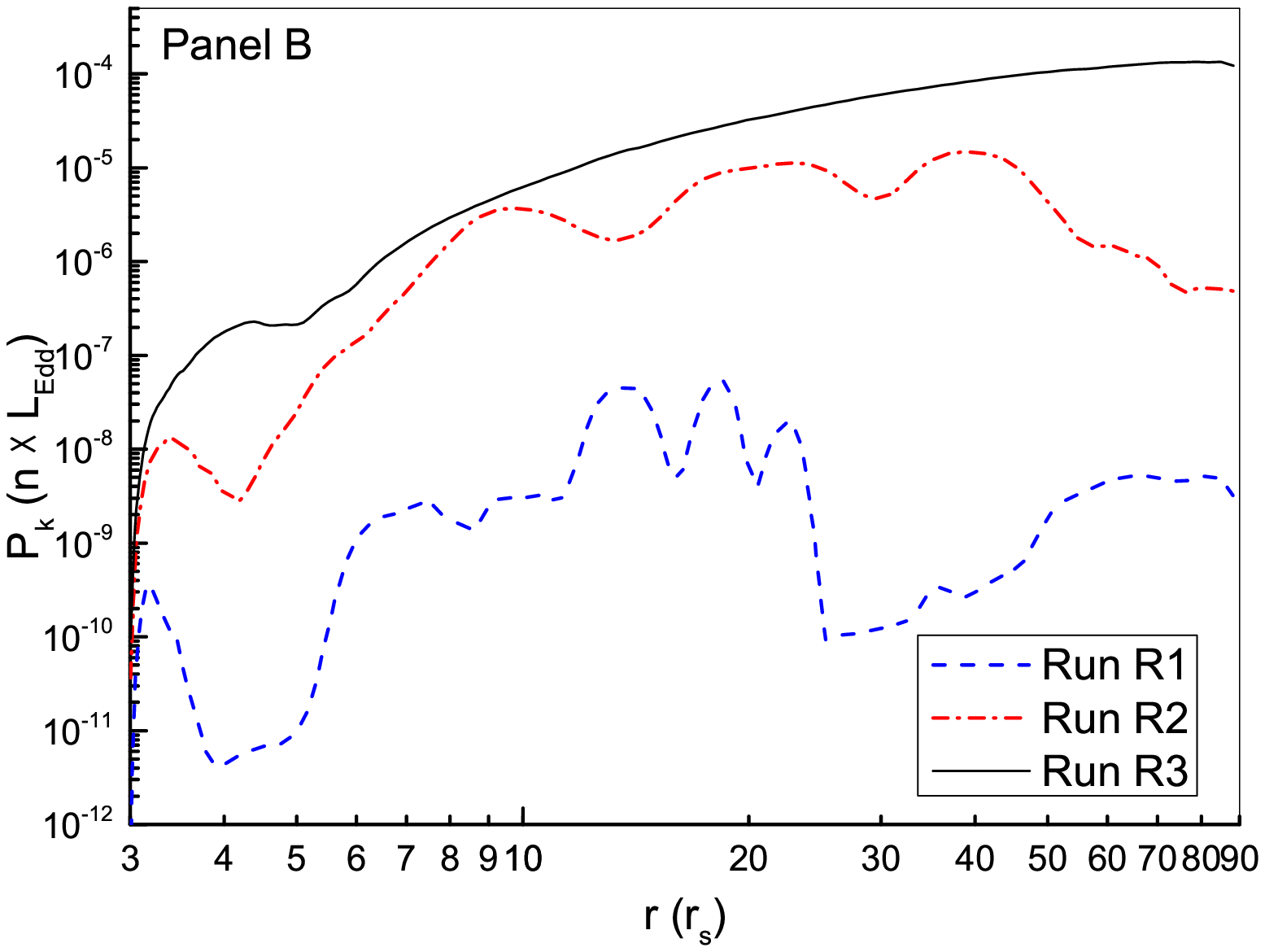}}}
\scalebox{0.35}[0.35]{\rotatebox{0}{\includegraphics[bb=45 20 500 360]{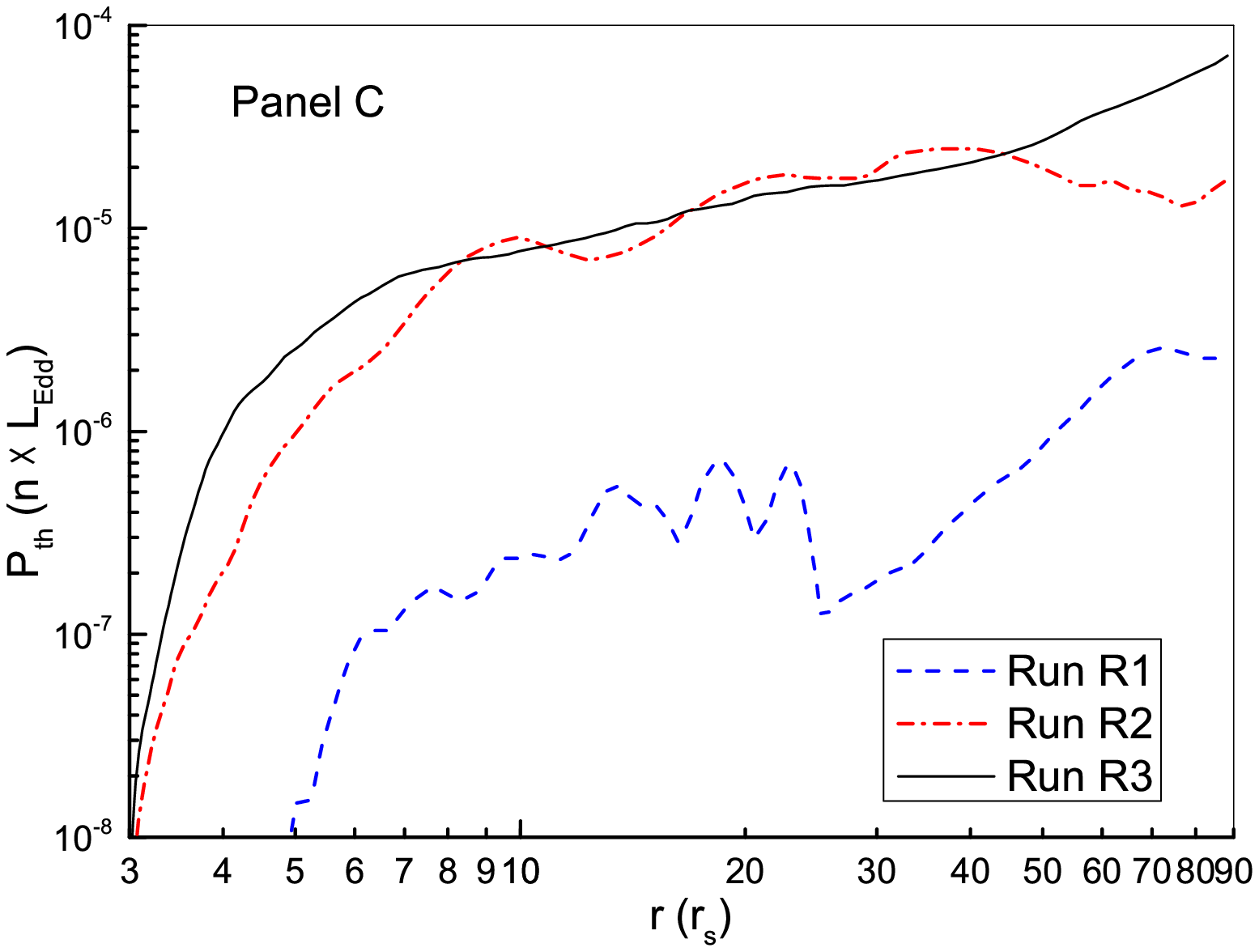}}}

\ \centering \caption{Time-averaged quantities as functions of $r$. panel (A): wind mass flux; panel (B): kinetic power of wind; panel (C): thermal energy flux carried by wind. The letter ``n'' is defined in Equation (10).}
\label{fig 7}
\end{figure*}

To quantitatively study the angular distribution of the physical variable of the irradiated gas, we plot Figure \ref{fig 5}. Figure \ref{fig 5} shows the angular profiles of the time-averaged variables at the outer boundary. We time-average 100 output files over the time range of 1.0-2.0 $T_{\rm orb}$ to obtain time-averaged values of the variables. The variables are radial column density (panel A), radial velocity (panel B), Mach number in the poloidal direction (panel C), mass flux density (panel D), and accumulated mass outflow rate (panel E). We define the accumulated mass outflow rate as $\Delta \dot{m}(\theta)=4\pi r^2_{out} \int_{\rm 0^\circ}^{\theta}\rho \max (v_r, 0) \sin\theta d\theta$. The angular distribution of radial column density strongly depends on the disk luminosity. In panel (A), the vertical axis is in units of $\rho_0l_0$. There is “$n$” in the density units. Therefore, $\rho_0l_0$ can be any value because $n$ can change so that the optically thin assumption is applicable. The column density (in units of $\rho_0l_0$) reaches 30 at the equator. This does not mean that the real optical depth is larger than 1. As shown in panel (A), the more luminous the disk, the slower the decrease of column density from the equator to the pole axis. Due to the increase of disk luminosity,  the radiation force becomes stronger and then the corona becomes thicker. For Run R1 with $\Gamma=0.25$, column density decreases three orders of magnitude from $\theta =90^{o}$ to $\theta =58^{o}$. At $\theta =\sim55^{o}$, the radial velocity of outflows reaches maximum, i.e. about 930 Km s$^{-1}$, which is much smaller than sound speed. Within the angular range of $\theta =70^{o}$--$90^{o}$, the radial velocity of outflows is smaller than 170 Km s$^{-1}$, while $98.7\%$ of wind mass flux is from the region $\theta =60^{o}$-- $90^{o}$. In fact, run R1 has vary low wind mass flux. For run R2 with $\Gamma=0.5$, column density decreases three orders of magnitude from  $\theta =90^{o}$ to $36^{o}$. The supersonic outflows are launched in the angle range $\theta=20^{o}$-- $34^{o}$ and their maximum radial velocity is about $1.0\times10^4$ Km s$^{-1}$. However, $99.6\%$ of wind mass flux is from the angular range of $\theta =60^{o}$-- $90^{o}$. For run R3 with $\Gamma=0.75$, column density decreases three orders of magnitude from $\theta =90^{o}$ to $31^{o}$. The supersonic outflows are launched in the angle range $\theta=17^{o}$-- $30^{o}$ and $50^{o}$-- $80^{o}$ and their maximum radial velocity is about $1.2\times10^4$ Km s$^{-1}$ at $\theta=\sim60^{o}$. $97.0\%$ of wind mass flux is from the angular range of $\theta =60^{o}$-- $90^{o}$.

Figure \ref{fig 6_R3} presents the radial dependence of the Bernoulli parameter ($\text{Be}$) of outflows, for run R3. The Bernoulli parameter is defined as $\text{Be}\equiv \textbf{v}^2/2+\gamma P/(\gamma-1)\rho-GM_{\rm bh}/r$). We time-average the Bernoulli parameter on the time range of 1.0--2.0$T_{\rm orb}$. In Figure \ref{fig 6_R3}, the solid line is obtained by averaging over an angle between $\theta=10^{o}$ to $\theta=46^{o}$ and the dashed line is obtained by averaging over an angle between $\theta=46^{o}$ to $\theta=90^{o}$. Comparing two lines, the outflows at low latitude have slightly larger Bernoulli parameter than that at high latitude. For run R3, the Bernoulli parameter is below zero, but the Bernoulli parameter keeps increasing outward. This means that the radiation force keeps accelerating the winds when winds move from small to large radii. We can expect that due to the persistent acceleration by radiation force, the Bernoulli parameter of wind will become positive at some larger radii. Winds will escape from the BH gravitation potential. The winds can go to the galaxy scale to interact with the interstellar medium (Ciotti \& Ostriker 2007; Ciotti et al. 2017).

We further quantitatively study the radial dependence of outflow properties. We calculate the dependence of mass outflow rate on the radius as follows:
\begin{equation}
\dot {M}_{\rm out} (r)=4\pi r^2 \int_{\rm 0^\circ}^{\rm 90^\circ}
\rho \max (v_r, 0) \sin\theta d\theta,
\end{equation}
We also calculate the kinetic ($P_{\rm k}$) and thermal energy ($P_{\rm th}$) carried by the outflow
as follows:
\begin{equation}
P_{\rm k} (r)=2\pi r^2 \int_{\rm 0^\circ}^{\rm 90^\circ} \rho
\max(v_r^3,0) \sin\theta d\theta
\end{equation}
\begin{equation}
P_{\rm th} (r)=4\pi r^2 \int_{\rm 0^\circ}^{\rm 90^\circ} e
\max(v_r,0) \sin\theta d\theta
\end{equation}
These quantities are shown in Figure \ref{fig 7} for the models listed in Table 1. We can see that the mass outflow rate increases outward for all the models. We can use $\dot{M}_{\rm out} \propto r^\alpha$ to describe the radial distribution of mass outflow rate outside 6 $r_{s}$. Here, $\alpha = 1.48$, 1.56, and 1.95 correspond to runs R1, R2, and R3, respectively. On average, $\alpha=1.66$. This suggests that outflows can be launched at any radii. The mass flux at any radii includes the mass flux of outflow launched locally and the mass flux of outflow from smaller radii. For run R1, the kinetic energy carried by the outflow complexly varies with radius. In run R1, the outflow is very weak. Most of the outflow mass flux is in the angle range $\theta=60^{o}$-- $90^{o}$, where the velocity field inside $50 r_{\rm s}$ (see left panel in Figure \ref{fig 4}) is very complex and irregularly changes with time in small spatial scale. Therefore, we do not take run R1 as an example to study the outflow properties. For run R2, the kinetic energy carried by the outflow also increases outward until 40$r_{s}$ and then decreases outward. This suggests that the outflow forming at the large radius has low radial velocity, which is caused by the distribution of radial radiation force. The angular range of $\theta =70^{o}$-- $90^{o}$ (see panels D and Panel E in Figure \ref{fig 5}) contributes to the vast majority of mass outflow rate. Within this angular range, the ratio of the radial radiation force to the BH gravity at the large radius (i.e. 90$r_{s}$) is smaller than that at the small radius (i.e. 20 $r_{s}$) (see Figure \ref{fig 1}). Within the angular range of $\theta<70^{o}$, the ratio of the radial radiation force to the BH gravity at the large radius (i.e. 90$r_{s}$) is higher than that at small radius (i.e. 20 $r_{s}$). Therefore, at high latitude, fast-speed outflows form at large radius, while the high-latitude outflows slightly contribute to the mass outflow rate. For run R3, the kinetic energy carried by the outflow also increases outward until $75 r_{\rm s}$. Because of the increase of disk luminosity, the outflows launched at the large radii have higher outflow velocity, compared with the case of run R2.  In run R3, the thermal energy carried by the outflow increases outward from the inner boundary to the outer boundary.

\begin{figure}

\scalebox{0.5}[0.5]{\rotatebox{0}{\includegraphics[bb=45 20 500 360]{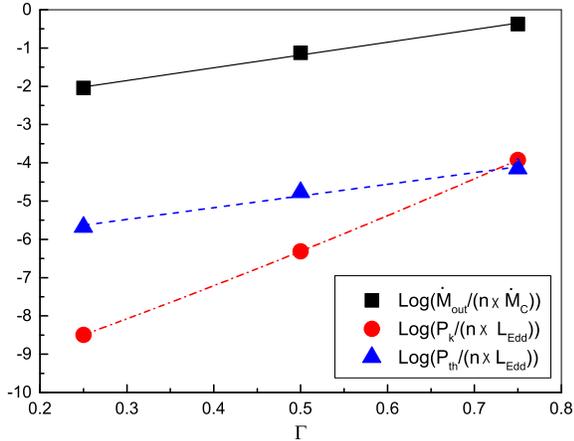}}}
 \ \centering \caption{Dependence of $\dot{M}_{\rm out}$, $P_{\rm k}$ and $P_{\rm th}$ on $\Gamma$. Data in this figure come from runs R1, R2, and R3 in Table 1. The letter ``n'' is defined in Equation (10).}
 \label{fig 8}
\end{figure}

Using data in Table 1, we plot Figure \ref{fig 8} to show dependence of $\dot{M}_{\rm out}$, $P_{\rm k}$, and $P_{\rm th}$ on $\Gamma$. We fit the data and obtain three relationships as follows: ${\dot{M}_{\rm out}}\propto 10^{3.3 \Gamma}$, ${P_{\rm k}}\propto 10^{9.1 \Gamma}$ and ${P_{\rm th}}\propto 10^{3.1 \Gamma}$.

\subsection{Dependence on temperature of corona}

\begin{figure}

\scalebox{0.5}[0.64]{\rotatebox{0}{\includegraphics[bb=80 360 520 700]{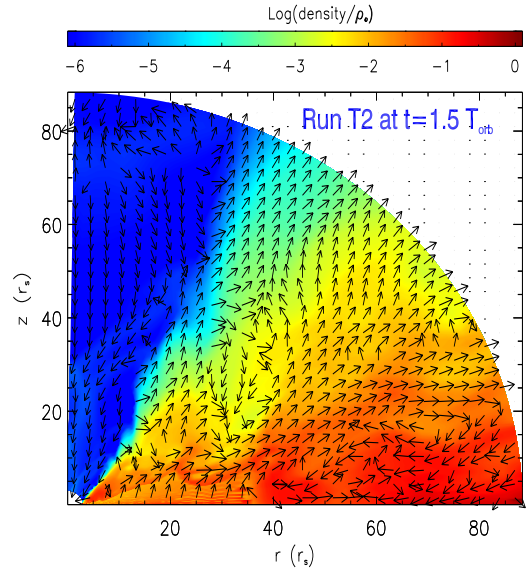}}}

\ \centering \caption{Snapshot of run T2 at $t=1.5 T_{\rm orb}$. Color denotes the logarithm density and arrows denote the direction of the poloidal velocity vector. }
\label{fig 9}
\end{figure}

\begin{figure}
\scalebox{0.40}[0.35]{\rotatebox{0}{\includegraphics[bb=10 20 500 360]{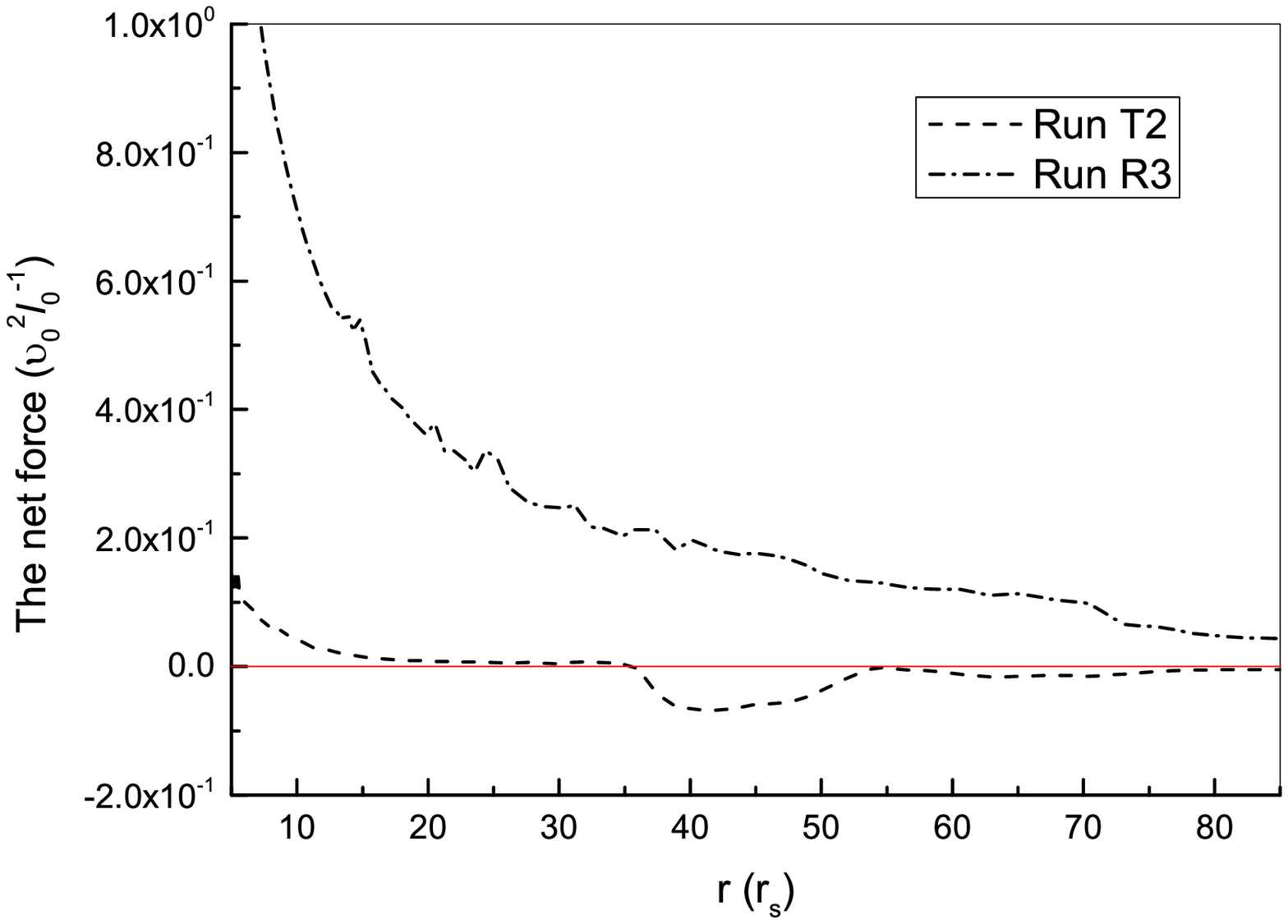}}}
\scalebox{0.40}[0.35]{\rotatebox{0}{\includegraphics[bb=10 20 500 360]{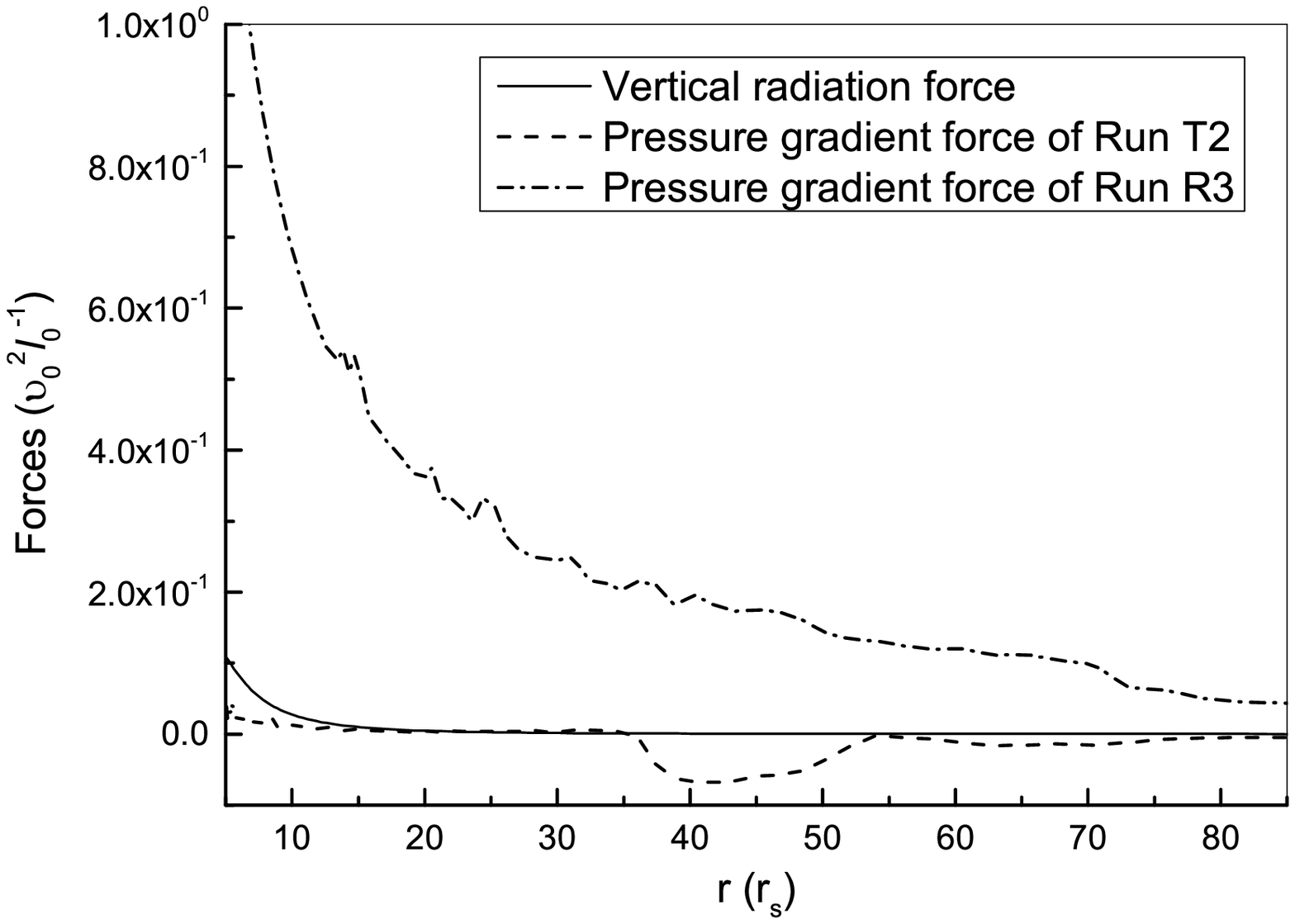}}}

 \ \centering \caption{Radial dependence of vertical forces at $\theta=89^{o}.964$ when $t=1.5T_{\rm orb}$, for runs R3 and T2. Top panel: the net vertical force (radiation force $+$ gravity $+$ pressure gradient force). In the top panel, the dotted-dashed line is for run R3 and the dashed line is for run T2. Bottom panel: the vertical radiation force (solid line), the vertical pressure gradient force of run T2 (dashed line), and the vertical pressure gradient force of run R3 (dotted-dashed line).}
 \label{fig 10}
\end{figure}

\begin{figure}
\scalebox{0.40}[0.35]{\rotatebox{0}{\includegraphics[bb=0 0 500 230]{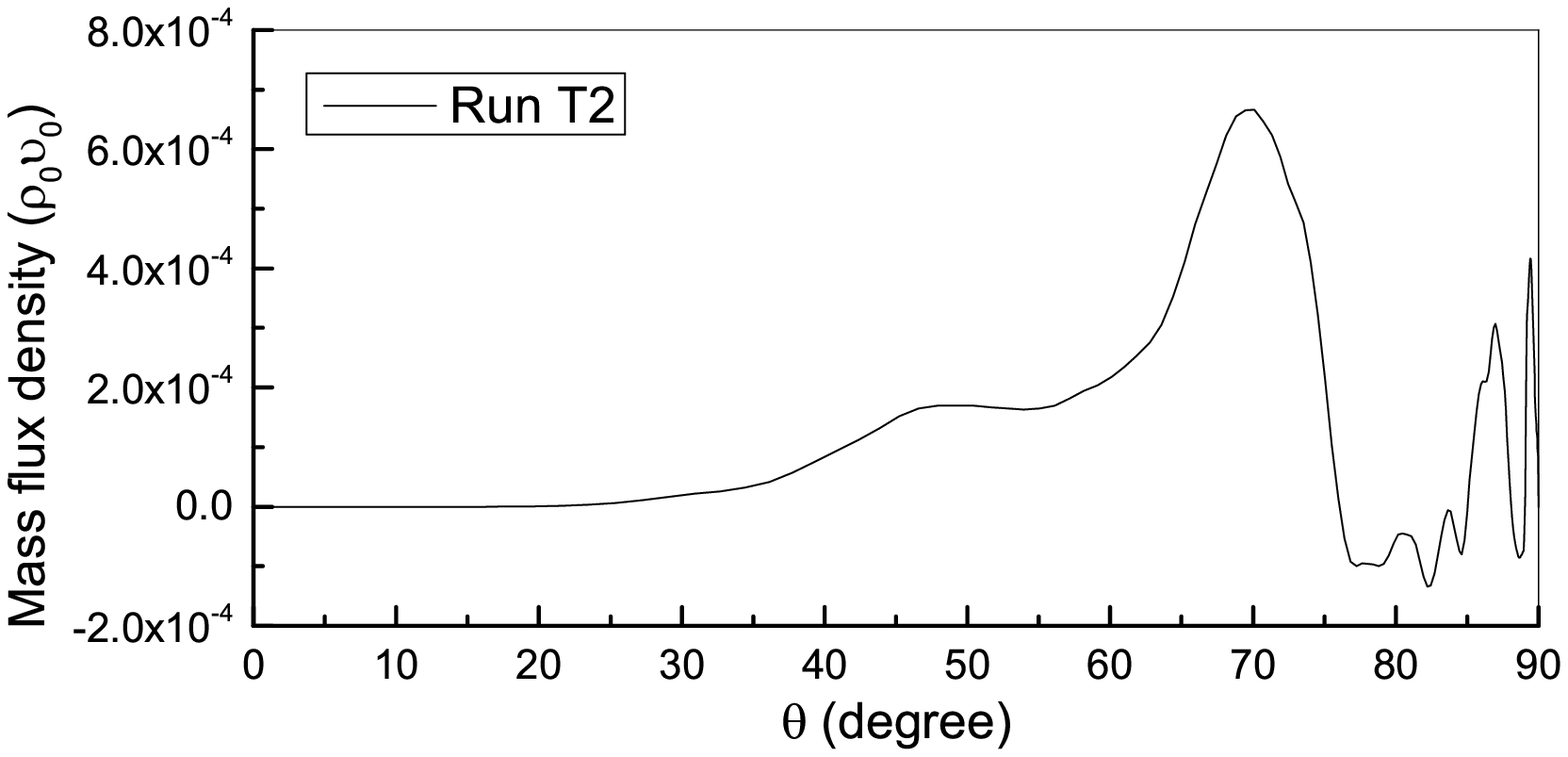}}}
\scalebox{0.40}[0.35]{\rotatebox{0}{\includegraphics[bb=0 0 500 230]{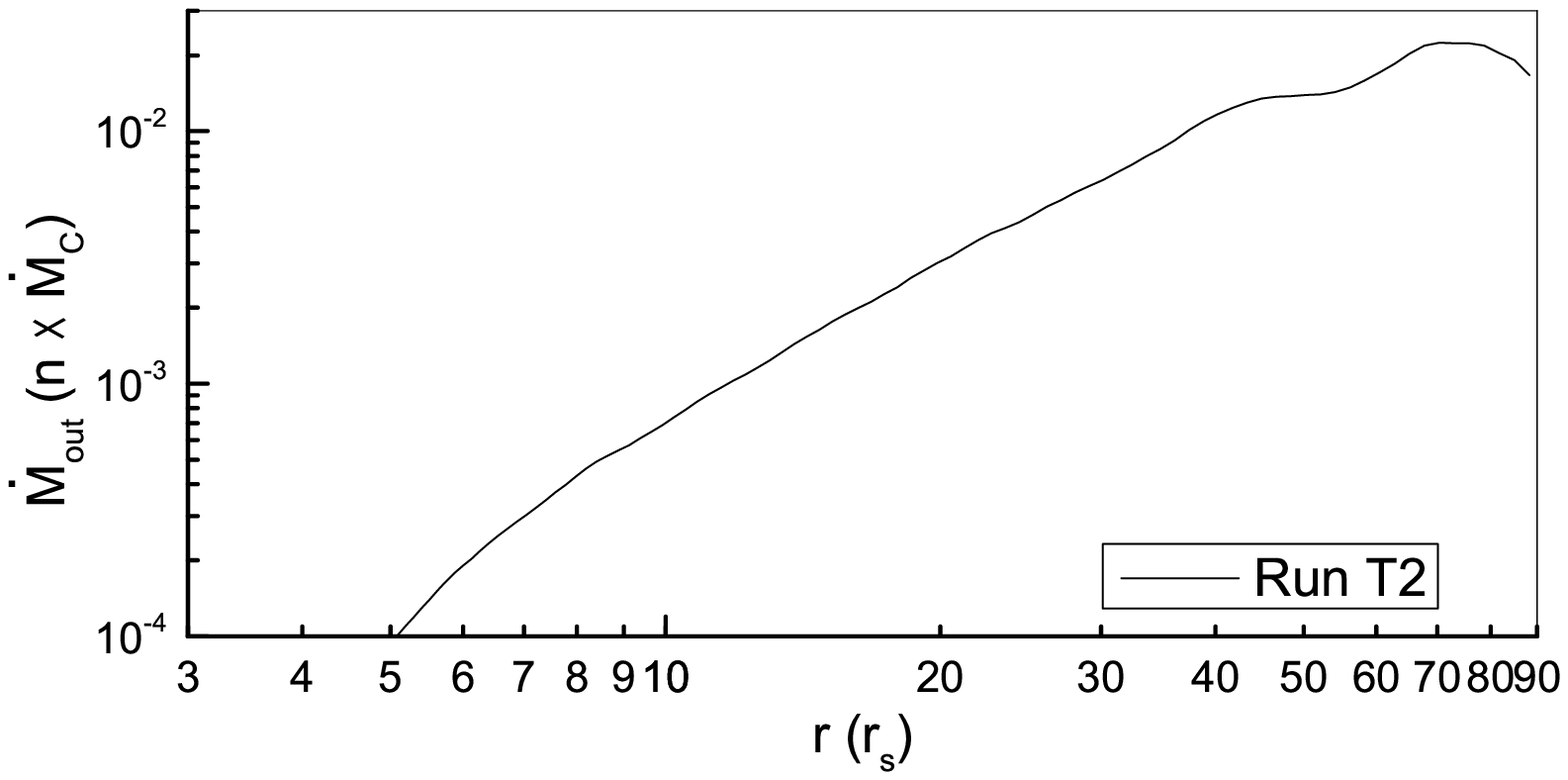}}}

 \ \centering \caption{Angular distribution (top panel) of mass flux density at the outer boundary and radial distribution (bottom panel) of mass outflow rate for run T2. The letter ``n'' in the bottom panel is defined in Equation (10).}
 \label{fig 11}
\end{figure}

\begin{figure}

\scalebox{0.4}[0.4]{\rotatebox{0}{\includegraphics[bb=45 20 500 360]{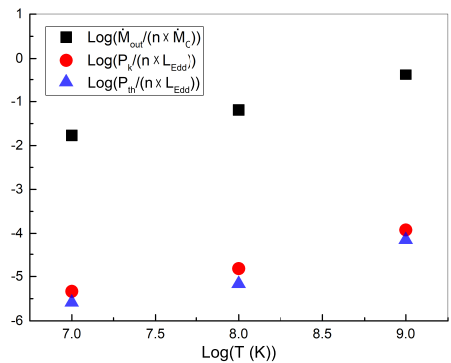}}}
 \ \centering \caption{Dependence of $\dot{M}_{\rm out}$, $P_{\rm k}$ and $P_{\rm th}$ at the outer boundary on corona temperature. Data in this figure come from Table 1. The letter ``n'' is defined in Equation (10).}
 \label{fig 12}
\end{figure}

Figure \ref{fig 9} shows the density snapshot of run T2 at $t=1.5T_{\rm orb}$, and arrows denote the direction of poloidal velocity. Comparing Figure \ref{fig 9} and the right panel of Figure 4, we find some differences. In run R3 with high corona temperature (i.e. $10^9$ K), gas is blown out almost at all the radii from the equator. In run T2 with low corona temperature (i.e. $10^7$ K), gas is blown out at the radius inside 35 $r_{\rm s}$. The $\theta$-component of radiation force at the plane of $\theta=\pi/2$ is an important factor to determine the vertical motion of gas at the equator. Figure \ref{fig 1} shows that the ratio of $\theta$-component radiation force to the BH gravity at the small radius (i.e. 20$r_{\rm s}$) is higher than that at the large radius (i.e. 90 $r_{\rm s}$). Therefore, when $\Gamma =0.75$, the gas at the equator can be blown out at the radius inside 35$r_{\rm s}$, even for the model with low corona temperature. At large radii (e.g. $>$35 $r_{\rm s}$), it is difficult for the $\theta$-component of radiation force to blow out the gas from the equator for the models with low corona temperature. In addition, Figure \ref{fig 10} shows radial dependence of vertical forces at $\theta=89^{o}.964$ when $t=1.5T_{\rm orb}$, for runs R3 and T2. At $\theta=89^{o}.964$, the vertical component of gravity is very small, compared with the vertical components of radiation force and pressure gradient force. For run R3, the direction of the net vertical force is upward in the region outside 4 $r_{\rm s}$; while for run T2, the direction of net vertical force is upward only within the 5-35 $r_{\rm s}$ radial range. Runs T2 and R3 have the same radiation force. In run T2, the vertical component of pressure gradient force is downward in the region outside 35 $r_{\rm s}$. Also, it is larger than the vertical component of radiation force. In run R3, the vertical component of pressure gradient force is upward. This is the reason why outflows can be launched from the equatorial plane at most of the radii in R3. Comparing runs T2 and R3, the strength of pressure gradient force is an important factor for driving gas away from the equator outside 35 $r_{\rm s}$, because the vertical radiation force is weak at $\theta=\pi/2$. In run T2, its initial temperature is set to be $10^7$ K and the temperature at the equator is kept to be $10^7$ K; while the gas above the equator has higher temperature than $10^7$ K due to the kinetic energy dissipation of turbulence or the compression work heating caused by the high-velocity outflow from the inner region. This is the reason why the vertical component of pressure gradient force is downward around the equator in run T2.

Figure \ref{fig 11} shows angular distribution (top panel) of mass flux density at the outer boundary and radial distribution (bottom panel) of mass outflow rate, for run T2. We can see that the outflows continuously take place in the angular range of $\theta=40^{o}$--75$^{o}$. In the angular range of $\theta=75^{o}$--90$^{o}$, the outflow intermittently takes place. As shown in Figure \ref{fig 9}, the inflow at the outer boundary takes place in the vicinity of $\theta=90^{o}$ when $t=1.5T_{\rm orb}$. The mass outflow rate of run T2 decreases outward outside 75$r_{\rm s}$. This is because that most of the outflow mass flux is in the angular range of $\theta=75^{o}$--90$^{o}$ and within the angular range the outflows forming inside 75$r_{\rm s}$ do not continuously move outward. As shown in Figure \ref{fig 9}, the inflows dominate outside 75$r_{\rm s}$ in the angle range $\theta=70^{o}$ --$90^{o}$. Therefore, the intermittent inflows at $\theta=75^{o}$--90$^{o}$ reduce the mass outflow rate outside 75$r_{\rm s}$.

Figure \ref{fig 12} shows dependence of $\dot{M}_{\rm out}$, $P_{\rm k}$, and $P_{\rm th}$ at the outer boundary on corona temperature. As mentioned above, Figure \ref{fig 12} clearly shows that$\dot{M}_{\rm out}$, $P_{\rm k}$, and $P_{\rm th}$ can be significantly reduced due to the decrease of corona temperature.

\section{Summary and discussion}
A hot corona commonly exists above the standard thin disk and so is irradiated by the disk. We perform two-dimensional hydrodynamical simulations to study the disk winds driven by radiation force from the corona. As a first step, we assume that there is a balance between heating and cooling in the irradiated corona. Only the compression work can heat the gas. We test effects of disk luminosity and corona temperature on the disk winds. Previous works found that the hot corona temperature is $\sim 10^9$K (e.g. Liu et al. 2003; Cao 2009; Liu et al. 2016). Therefore, we mainly focus on the properties of disk winds in the case of corona temperature $T_{\rm C}=10^9$K.

The disk luminosity is a key factor to determine the properties of disk winds. With the increase of disk luminosity, the disk winds get stronger and the angular range of producing disk winds becomes wider. We can use $\dot{M}_{\rm out}\propto 10^{3.3 L_{\rm D}/L_{Edd}}$ to describe the dependence of mass outflow rate at the outer boundary on the disk luminosity. In the case of low luminosity (e.g. 0.25$L_{\rm Edd}$), the disk winds are very weak and the velocity field is mussy at small spatial scales. In the case of high luminosity (e.g. 0.75$L_{\rm Edd}$), radiation force blows out gas from the equator and the blown-out gas moves upward to form large-scale circulation or outflows. The outflows exist in the angle range of $\theta=17^{0}$-- $90^{0}$. Their velocity can reach $1.0\times10^4$ Km s$^{-1}$ at $\theta=\sim60^{o}$. The supersonic outflows are launched at $\theta=\sim17^{o}$ --$30^{o}$ and $\theta=\sim50^{o}$ --$80^{o}$.

Corona temperature significantly influences the properties of outflows. In the case of low temperature (e.g. $T_{\rm C}=10^7$ K), gas pressure gradient force is weak; radiation force is not enough to blow the gas at large radius to form outflows.

Tajima \& Fukue (1996; 1998) used particle dynamics to study the disk winds driven by the radiation force. They found that the disk winds can be launched from the inner disk when $L_{\rm D}/L_{\rm Edd}>0.6$. If the radiation drag is taken into account, the disk winds can be launched when $L_{\rm D}/L_{\rm Edd}>0.8$. In our hydrodynamical simulations, we take into account the gas pressure gradient force. The pressure gradient force also plays an important role in driving the thin disk winds. Therefore, we find that the disk winds can be launched when disk luminosity is below $0.6 L_{\rm Edd}$.

We also note that (1) We set corona temperature at the equator to be a constant with radius. Our simulations indicate that the corona temperature is a key factor to determine the properties of outflows. Therefore, in the future, it is very necessary to study different radial temperature profile cases. (2) The heating and cooling are assumed to keep a balance in the corona. In the future, it is necessary to study corona wind with real heating and cooling processes. (3) A large-scale magnetic field can also certainly play an important role in driving outflows (Blandford \& Payne 1982). Both radiation force and magnetic field should be simultaneously considered in simulations. This work is in progress.

\acknowledgments{We thank the Fundamental Research Funds for the Central Universities (No. 2018CDXYWU0025). D. B. is supported in part by the National Program on Key Research and Development Project of China (Grant No. 2016YFA0400704),  the Natural Science Foundation of China (grants 11573051, 11633006, 11773053, and 11661161012), the Natural Science Foundation of Shanghai (grant 16ZR1442200), and the Key Research Program of Frontier Sciences of CAS (No. QYZDJSSW-SYS008). This work made use of the High Performance Computing Resource in the Core Facility for Advanced Research Computing at Shanghai Astronomical Observatory.}

\end{document}